\documentclass[%
reprint,
superscriptaddress,
 nofootinbib,
 amsmath,amssymb,
 aps,
 pra,
]{revtex4-2}

\pdfoutput=1

\usepackage[utf8]{inputenc}
\usepackage{CJKutf8}
\usepackage{amsthm}
\usepackage{amssymb}
\usepackage{mathrsfs}
\usepackage{physics}
\usepackage{bbm}
\usepackage{bm}
\usepackage{quantikz}
\usepackage[ruled,vlined]{algorithm2e}
\usepackage{graphicx}
\usepackage{dcolumn}
\usepackage{hyperref}
\usepackage{xcolor,soul}
\usepackage[english]{babel}
\usepackage{romannum}

\usepackage[boxsize=0.5em, centertableaux]{ytableau}

\makeatletter
\newcommand\RedeclareMathOperator{%
  \@ifstar{\def\rmo@s{m}\rmo@redeclare}{\def\rmo@s{o}\rmo@redeclare}%
}
\newcommand\rmo@redeclare[2]{%
  \begingroup \escapechar\m@ne\xdef\@gtempa{{\string#1}}\endgroup
  \expandafter\@ifundefined\@gtempa
     {\@latex@error{\noexpand#1undefined}\@ehc}%
     \relax
  \expandafter\rmo@declmathop\rmo@s{#1}{#2}}
\newcommand\rmo@declmathop[3]{%
  \DeclareRobustCommand{#2}{\qopname\newmcodes@#1{#3}}%
}
\@onlypreamble\RedeclareMathOperator
\makeatother

\DeclareMathOperator{\diag}{diag}

\RedeclareMathOperator{\ln}{ln}
\DeclareMathOperator{\polylog}{polylog}
\DeclareMathOperator{\SC}{SC}
\DeclareMathOperator{\MC}{MC}
\DeclareMathOperator{\TC}{TC}
\DeclareMathOperator{\F}{F}
\DeclareMathOperator{\T}{T}
\DeclareMathOperator*{\expect}{\mathbbm{E}}
\DeclareMathOperator*{\variance}{\mathbbm{V}}

\newcommand{\RNum}[1]{\uppercase\expandafter{\romannumeral #1\relax}}
\newcommand{\myquad}[1][1]{\hspace*{#1em}\ignorespaces}

\newtheorem{theorem}{Theorem}
\newtheorem{corollary}[theorem]{Corollary}
\newtheorem{lemma}[theorem]{Lemma}
\newtheorem{definition}[theorem]{Definition}

\def \d {\mathrm{d}}

\def \ran {\normalfont{\text{ran}}}

\def \Ex {\{E_x\}_{x\in\mathcal{X}}}
\def \Gi {\{G_i\}_{i\in[k]}}
\def \Fxi {\{F_{x}^{i}\}_{x\in\mathcal{X}_i}}

\def \id {\mathbbm{I}}

\begin{document}
\begin{CJK*}{UTF8}{gbsn}

\preprint{APS/123-QED}

\title{Sample Optimal and Memory Efficient Quantum State Tomography}

\author{Yanglin Hu (胡杨林)}
\email{yanglin.hu@u.nus.edu}
\affiliation{%
 Centre for Quantum Technologies, 
 National University of Singapore, 
 3 Science Drive 2, 117543 Singapore, Singapore
}%

\author{Enrique Cervero-Mart{\'i}n}
\email{enrique.cervero@u.nus.edu}
\affiliation{%
 Centre for Quantum Technologies, 
 National University of Singapore, 
 3 Science Drive 2, 117543 Singapore, Singapore
}%

\author{Elias Theil}
\email{edmt@math.ku.dk}
\affiliation{
Department of Mathematical Sciences, 
University of Copenhagen,
Universitetsparken 5, DK-2100 Copenhagen,Denmark
}

\author{Laura Mančinska}
\email{mancinska@math.ku.dk}
\affiliation{
Department of Mathematical Sciences, 
University of Copenhagen,
Universitetsparken 5, DK-2100 Copenhagen, Denmark
}

\author{Marco Tomamichel}%
\email{marco.tomamichel@nus.edu.sg}
\affiliation{%
 Centre for Quantum Technologies,
 National University of Singapore, 
 3 Science Drive 2, 117543 Singapore, Singapore
}%
\affiliation{%
 Department of Electrical and Computer Engineering,
 National University of Singapore,
 4 Engineering Drive 3, 117583 Singapore, Singapore
}%

\date{\today}

\begin{abstract}
    Quantum state tomography is the fundamental physical task of learning a complete classical description of an unknown state of a quantum system given coherent access to many identical samples of it. The complexity of this task is commonly characterised by its sample-complexity: the minimal number of samples needed to reach a certain target precision of the description.
    While the sample complexity of quantum state tomography has been well studied, the memory complexity has not been investigated in depth. Indeed, the bottleneck in the implementation of na\"ive sample-optimal quantum state tomography is its massive quantum memory requirements. In this work, we propose and analyse a quantum state tomography algorithm which retains sample-optimality but is also memory-efficient. Our work is built on a form of unitary Schur sampling and only requires streaming access to the samples. 
\end{abstract}

\maketitle
\end{CJK*}

\section{Introduction}

Experiments aiming to utilize quantum advantages in computation and communication require precise preparation, manipulation and measurements of quantum states. Quantum state tomography is an essential subroutine in order to characterise quantum states in each step of such experiments~\cite{Paris_2004,Lu_2016,Yu_2020,Barreiro_2011,Fluhmann_2020}. 
However, experimental realization of quantum state tomography is bottlenecked by the complexity of existing algorithms, especially when applied to large quantum systems~\cite{Banaszek_2013}. 
This motivates the development of efficient algorithms for quantum state tomography, the problem which we attempt to tackle in this work. In contrast to recent lines of work on shadow tomography~\cite{Aaronson_2020,Chen_2024b} and classical shadows~\cite{Huang_2020,Chen_2024b,Grier_2024a,Nguyen_2022,Grier_2024b}, we are here interested in learning a complete characterisation of the quantum state and not only its behaviour under a restricted set of measurements.

Formally, the goal of a quantum state tomography algorithm receiving $n$ copies of a $d$ dimensional quantum state $\rho$ is to produce an estimate $\hat{\rho}$ of $\rho$ that satisfies the \emph{probably approximately correct} condition (PAC condition for short), i.e. $\Pr[\Delta(\rho,\hat{\rho})\leq \epsilon]\geq 1-\xi$, where $\Delta$ is a chosen accuracy measure, $\epsilon$ is the target accuracy and $\xi$ is the upper bound on the failure probability. Note that quantum state tomography is quintessentially probabilistic due to the randomness introduced by Born's rule, and, as a consequence, we ought not hope for a deterministic guarantee that our estimate is accurate.
However, the failure probability $\xi$ can be exponentially suppressed in the number of repetitions of the tomography algorithm (see Appendix~\ref{apd:cond_equiv}), and we thus fix $\xi$ to be constant for the following discussions.

The three factors of interest in the development of efficient tomography algorithms are the {sample complexity}, the {memory complexity} and the {time complexity}. 
Suppose a tomography algorithm achieves accuracy $\epsilon$ in the accuracy measure $\Delta$, we are interested in the following properties:
\begin{itemize}
    \item The \emph{sample complexity}, $\SC_{\Delta}(\epsilon,d,r)$, specifies the minimum required number of samples $n$ of the unknown input state $\rho$ to achieve target accuracy $\epsilon$. Here, $d$ is the dimension of the state space and $r\leq d$ is the rank (if known) of the unknown state. 
    \item The \emph{memory complexity}, $\MC_{\Delta}(\epsilon,d,r)$, or space complexity as in~\cite{Crouch_2016, LeGall_2006}, gives the maximum number of qubits that need to be stored through an execution of the tomography algorithm at any given time. 
    The memory complexity therefore includes both the number of input states operated on at any time as well as any required auxiliary working registers.
    \item Lastly, the \emph{time complexity}, $\TC_{\Delta}(\epsilon,d,r)$, gives the number of basic quantum operations needed in an instance of the tomography algorithm. 
\end{itemize}
Note that in the following, \emph{memory complexity} refers only to \emph{quantum} memory and is unrelated to any additional classical memory that needs to be maintained throughout an execution of our algorithms. 
This distinction is justified since quantum memory is the  more expensive resource.
For brevity, we will also omit the dependence on $(\epsilon, d, r)$ in the following when it is clear from context.

An important fact is that memory complexity and sample complexity need not be equivalent.
This is especially the case when the input copies of the unknown state are presented in a sequential manner, in which case the memory complexity can be significantly lower than the sample complexity by wisely processing the sequential inputs\,---\,as is usually observed in streaming/online algorithms\,---\,rather than storing all copies and performing a coherent operation at the end.

Intuitively, the sample complexity, the memory complexity and the time complexity should all depend on the dimension $d$ of the quantum states, their rank $r$, as well as the desired accuracy $\epsilon$ and the choice of accuracy measure $\Delta$. In particular, in this work we focus on the accuracy measures of infidelity and trace distance defined in Eqs.~\eqref{eq:def_trace_distance} and~\eqref{eq:def_infidelity}, which we denote by $\Delta=\F$ and $\Delta=\T$ respectively. We will by default use $\delta$ and $\epsilon$ for accuracies in infidelity and trace distance, respectively.

Different tomography algorithms exist which analyse the sample complexity when allowing single qudit measurements~\cite{Kueng_2017,Guta_2020}, fully coherent measurements~\cite{ODonnell_2016,ODonnell_2017,Haah_2017} or $t$-coherent measurements (measurements on a subset of $t$ input qudits)~\cite{Chen_2024a}. These algorithms, due to their respective access model, give us bounds on their memory complexity. However, when moving away from fully coherent access, these algorithms are not necessarily sample-optimal. 

In this work we aim for a more principled approach: our target is to minimize the memory complexity for a sample-optimal algorithm. 
We leverage the sample-optimal algorithms of~\cite{ODonnell_2016,ODonnell_2017,Haah_2017} with the recent streaming and memory-efficient algorithm for unitary Schur sampling of~\cite{Cervero_2024} to provide a streaming, sample-optimal and memory-efficient algorithm for quantum state tomography. It achieves the optimal sample complexities 
\begin{align}
    \SC_{\F}=O\bigg(\frac{dr}{\delta}\ln\frac{d}{\delta}\bigg) \quad \textrm{and} 
    \quad \SC_{\T}=O\bigg(\frac{dr}{\epsilon^2}\bigg), \label{eq:our-samp-opt}
\end{align}
whilst also improving the memory complexities to
\begin{align}
    \MC_{\F}=O\bigg(dr\ln\frac{d}{\delta}\bigg) \quad \textrm{and} \quad
    \MC_{\T}=O\bigg(dr\ln\frac{d}{\epsilon}\bigg), \label{eq:our-mem-efficient}
\end{align}
respectively.
The other improvement between our tomography algorithm (presented in Algorithm~\ref{alg:stream_tomo}) and those of~\cite{ODonnell_2016,ODonnell_2017,Haah_2017} is the discretization of the continuous measurement over the Haar measure, which allows for a memory-efficient implementation at the cost of an exponentially suppressible failure probability. This constitutes the main technical contribution of our work. We do not optimize the time complexity (it is exponential in some parameters), leaving this for future work.

The remainder of the paper is organized as follows. In Section~\ref{sec:comparison}, we review existing tomography algorithms, summarize their sample and memory complexities in Table~\ref{tab:summary}, compare them to our algorithm and highlight the advantage of our work in Tables~\ref{tab:comparison_fullycoherent} and~\ref{tab:comparison_tcoherent}. We introduce basic preliminaries such as distance measures, Young diagrams and Schur-Weyl duality in Section~\ref{sec:preliminaries}. Section~\ref{sec:povm_construction} contains the main technical contribution of our work, which constructs a memory-efficient discretization of the continuous Haar random measurement. Using this discrete measurement, we present our algorithm and prove its sample optimality and memory efficiency for infidelity and trace distance in Section~\ref{sec:algorithm}. Finally in Section~\ref{sec:discussion}, we discuss the possibility of improving the time complexity, as well as the equivalence between different accuracy conditions.

\section{Comparison with previous work}
\label{sec:comparison}
\begin{table*}[htbp!]
\begin{center}
    \begin{tabular}{ |c|c|c|c|c|c|c|} 
    \hline
    \rule{0pt}{3ex} Access model & $\SC_{\F}$ & $\MC_{\F}$ & $\SC_{\T}$ & $\MC_{\T}$  \\ [0.5ex] 
    \hline
    \rule{0pt}{3ex} Fully coherent (rank $r$) & $O\big(\frac{dr}{\delta} \ln\frac{d}{\delta}\big)$ & $O\big(\frac{dr}{\delta} \ln\frac{d}{\delta}\ln d\big)$ & $O\big(\frac{dr}{\epsilon^2}\big)$ & $O\big(\frac{dr}{\epsilon^2}\ln d\big)$ \\  [0.5ex]
    \hline
    \rule{0pt}{3ex}  Fully coherent lower bound (rank $r$) & $\Omega\big(\frac{dr}{\delta }\big)$ &    & $\Omega\big(\frac{dr}{\epsilon^2\ln d/(r\epsilon)}\big)$ ($\Omega\big(\frac{dr}{\epsilon^2}\big)$ if $r=d$) &    \\ [0.5ex]
    \hline
    \rule{0pt}{3ex} Single-copy (rank $r$)  &  $O\big(\frac{dr^2}{\delta}\ln\frac{r}{\delta}\big)$  & $O(\ln d)$  &  $O\big(\frac{dr^2}{\epsilon^2}\ln d\big)$ & $O(\ln d)$ \\ [0.5ex]
    \hline
    \rule{0pt}{3ex} Single-copy lower bound (rank $d$)  &  $\Omega\big(\frac{d^3}{\delta}\big) $  &     &  $\Omega\big(\frac{d^3}{\epsilon^2}\big)$ &   \\ [0.5ex]
    \hline
    \rule{0pt}{3ex} $t$-coherent (rank $r$)  &    &    & $O\big(\frac{d^3}{\sqrt{t} \epsilon^2}(\ln t)^3\big)$   & $O(t\ln d)$ \\ [0.5ex]
    \hline
    \rule{0pt}{3ex} $t$-coherent lower bound (rank $d$)  &    &    & $\Omega\big(\frac{d^3}{\sqrt{t} \epsilon^2}\big)$   &  \\ [0.5ex]
    \hline
    \rule{0pt}{3ex} $d^2$-coherent (rank $r$)  &    &    & $O\big(\frac{d^2}{ \epsilon^2}(\ln d)^3\big)$   & $O(d^2\ln d)$ \\ [0.5ex] 
    \hline
    \rule{0pt}{3ex} $d^2$-coherent lower bound (rank $d$)  &    &    & $\Omega\big(\frac{d^2}{\epsilon^2}\big)$   & \\ [0.5ex]
    \hline
    \rule{0pt}{3ex} \textbf{Streaming fully coherent (rank $r$) (this work)}  & $O\big(\frac{dr}{\delta} \ln\frac{d}{\delta}\big)$ & $O\big(dr\ln\frac{d}{\delta} \big)$ & $O\big(\frac{dr}{\epsilon^2} \big)$ & $O\big(dr\ln\frac{d}{\epsilon}\big)$ \\ [0.5ex]
    \hline
\end{tabular}
\end{center}
\caption{Comparison between previous algorithms and ours. Here $\F$ denotes infidelity, and $\T$ denotes trace distance. For the references and how we unify the references to obtain this table, see Section~\ref{sec:comparison}. }
\label{tab:summary}
\end{table*}

\begin{table*}[htbp!]
\begin{center}
    \begin{tabular}{ |c|c|c|c|c|} 
    \hline
    \rule{0pt}{3ex} Merit  & This work & Fully coherent & Our adv. & When we have advantage \\ [0.5ex]
    \hline
    \rule{0pt}{3ex} $\SC_{\F}$ &  $O\big(\frac{dr}{\delta}\ln\frac{d}{\delta}\big)$  & $O\big(\frac{dr}{\delta}\ln\frac{d}{\delta}\big)$ & None & None \\ [0.5ex]
    \hline
    \rule{0pt}{3ex} $\MC_{F}$ & $O\big(dr\ln \frac{d}{\delta}\big)$  & $O\big(dr\textcolor{red}{\frac{\ln d}{\delta}}\ln\frac{d}{\delta}\big)$ & $O\big(\frac{\ln d}{\delta}\big)$ & For large $d$ or small $\delta$ \\ [0.5ex]
    \hline
    \rule{0pt}{3ex} $\SC_{\T}$ &  $O\big(\frac{dr}{\epsilon^2}\big)$  & $O\big(\frac{dr}{\epsilon^2}\big)$ & None & None \\ [0.5ex]
    \hline
    \rule{0pt}{3ex} $\MC_{\T}$ & $O\big(dr\textcolor{blue}{\ln\frac{d}{\epsilon}}\big)$  &  $O\big(dr\textcolor{red}{\frac{\ln d}{\epsilon^2}}\big)$  &  $O\big(\frac{\ln d}{\epsilon^2\ln d/\epsilon}\big)$ &  For large $d$ or small $\delta$ \\  [0.5ex]
    \hline
\end{tabular}
\end{center}
\caption{Comparison between ours and the fully coherent algorithm, obtained from Table~\ref{tab:summary}. }
\label{tab:comparison_fullycoherent}
\end{table*}

\begin{table*}[htbp!]
\begin{center}
    \begin{tabular}{ |c|c|c|c|c|} 
    \hline
    \rule{0pt}{3ex} Merit  & This work & $d^2$-coherent & Our adv. & When we have advantage. \\ [0.5ex]
    \hline
    \rule{0pt}{3ex} $\SC_{\T}$ &  $O\big(\frac{dr}{\epsilon^2}\big)$  & $O\big(\frac{d^2}{\epsilon^2}\textcolor{red}{(\ln d)^3}\big)$ & $O\big(\frac{d}{r}(\ln d)^3\big)$ & For large $d$ or small $r$ \\ [0.5ex]
    \hline
    \rule{0pt}{3ex} $\MC_{\T}$ & $O\big(d\textcolor{blue}{r\ln\frac{d}{\epsilon}}\big)$  &  $O\big(d\textcolor{red}{d\ln d}\big)$  &  $O\big(\frac{d\ln d}{r\ln d/\epsilon}\big)$ & For large $d$ or small $r$\\  [0.5ex]
    \hline
\end{tabular}
\end{center}
\caption{Comparison between ours and the $d^2$-coherent algorithm, obtained from Table~\ref{tab:summary}. Note that we have a logarithmic disadvantage when $\epsilon$ is very small. For a more thorough comparison between ours and a family of $t$-coherent algorithm with a sample complexity of $O\big(\frac{d^2}{\epsilon^2}\polylog(d)\big)$, see Section~\ref{sec:tcoh}.}
\label{tab:comparison_tcoherent}
\end{table*}

Quantum state tomography has been intensively investigated, reaching several milestones such as~\cite{Haah_2017,ODonnell_2016,ODonnell_2017,Guta_2020,Chen_2023,Chen_2024a}. However, the notation in these papers is not wholly unified: some use the PAC condition whilst others use the \emph{second moment of distance} (SMD) condition (see Appendix~\ref{apd:cond_equiv} for details). Further, some works keep all logarithmic factors while some omit all of them. 
In this article, we unify all results under the more frequently used PAC condition via Lemma~\ref{lem:cond_equiv} which converts the SMD condition into the PAC condition at the cost a logarithmic factor $O(\ln\frac{1}{\xi})$, where $\xi$ is the failure probability in the PAC condition. 
In addition, we keep all $O(\ln d)$ and $O(\ln \frac{1}{\epsilon})$ factors explicitly, but treat the failure probability $\xi$ in the PAC condition as a constant and as such omit all $O(\ln\frac{1}{\xi})$ factors. This is partly due to the fact that most previous works set $\xi$ to be a constant and absorb it into the big $O$ notation. 
Lemma~\ref{lem:pac_repetition} further justifies this choice, as it states that $\xi$ can be suppressed exponentially with enough repetitions of the tomography algorithm. 

As previously mentioned, the sample complexity of a quantum state tomography algorithm depends on the dimension and rank of the input states as well as the accuracy and accuracy measure. In this section, we will focus on learning a $d$-dimensional $r$-rank quantum state $\rho$.
We concentrate on upper bounds and lower bounds on $\SC_{\F}$ and $\SC_{\T}$. It is worth noting that for some nondecreasing function $f$, an algorithm with 
\begin{align}
    \SC_{\F}=O\big(f(\delta,d,r)\big), 
\end{align}
is also an algorithm with
\begin{align}
    \SC_{\T}=O\big(f(\epsilon^2,d,r)\big).
\end{align}
This is due to the Fuchs-van de Graaf inequality $T(\rho,\hat{\rho})\leq \sqrt{2I(\rho,\hat{\rho})}$ where $T(\,\cdot\,,\,\cdot\,)$ and $I(\,\cdot\,,\,\cdot\,)$ are the trace distance and the infidelity, respectively. An algorithm achieving $I(\rho,\hat{\rho})\leq \delta$ therefore also achieves $T(\rho,\hat{\rho})\leq \sqrt{2\delta}$. Conversely, a lower bound 
\begin{align}
    \SC_{\T}=\Omega\big(f(\epsilon^2,d,r)\big),
\end{align}
also implies a lower bound
\begin{align}
    \SC_{\F}=\Omega\big(f(\delta,d,r)\big). 
\end{align}

In fact, the sample complexity also depends on whether the algorithm allows for coherent measurements on many copies of the input state or only on single copy measurements. We discuss the sample complexities existing in the current literature in each of the aforementioned variations in the following subsections and present a concise comparison in Table \ref{tab:summary}.

\subsection{Fully coherent strategies}

In~\cite{Haah_2017}, the authors proposed a fully coherent tomography algorithm with 
\begin{align}
    \SC_{\F} = O\left(\frac{dr}{\delta}\ln \frac{d}{\delta}\right), 
\end{align}
as can be seen from~\cite[Eq.~(15)]{Haah_2017}. This immediately gives $\SC_{\T} = O\big(\frac{dr}{\epsilon^2}\ln \frac{d}{\epsilon}\big)$ derived from $\SC_{\F}$ by the Fuchs-van de Graaf inequality. Here, we have absorbed an additive logarithmic factor $\frac{1}{\delta}\ln\frac{1}{\xi}$ into the big O notation. 
Concurrently,~\cite[Corollary~1.4]{ODonnell_2016} proposed a slightly different fully coherent tomography algorithm with 
\begin{align}
    \SC_{\T} = O\left(\frac{dr}{\epsilon^2}\right), 
\end{align}
providing a logarithmic improvement over \cite{Haah_2017}. 
Later, in~\cite{ODonnell_2016b}, the authors proved that the algorithm in~\cite{Haah_2017} also achieves $\SC_{\T} = O\big(\frac{dr}{\epsilon^2}\big)$, removing the logarithmic factor.\footnote{It is worth noting that the original proof in~\cite{ODonnell_2016} was given for the SMD condition (see~\cite[Section~4]{ODonnell_2016}), but is converted to the PAC condition in the subsequent~\cite{ODonnell_2016b} at the cost of a logarithmic overhead $O\big(\ln\frac{1}{\xi}\big)$, as required by Lemma~\ref{lem:cond_equiv} in Appendix~\ref{apd:cond_equiv}.}

As for lower bounds for fully coherent tomography algorithms: 
In~\cite[Theorem~3]{Haah_2017} it is shown that when learning in trace distance
\begin{align}
    \SC_{\T}=\Omega\left(\frac{dr}{\epsilon^2\ln \frac{d}{r\epsilon}}\right), 
\end{align}
and that when $r=d$, $\SC_{\T}=\Omega\big(\frac{d^2}{\epsilon^2}\big)$, matching the upper bound in~\cite{Haah_2017,ODonnell_2016} without any logarithmic factors. 
Additionally, in~\cite{Yuen_2023}, the lower bound for the infidelity was improved to 
\begin{align}
    \SC_{\F}=\Omega\left(\frac{dr}{\delta}\right).
\end{align}

With respect to the memory complexity, in a naïve application of a fully coherent tomography algorithm, one simply stores each sample received and perform a fully coherent measurement on them in the end when all samples received. Noting that each qudit is stored on $O(\ln d)$ qubits, the memory complexity (neglecting any auxiliary qubits required to perform a continuous measurement over the Haar measure) is the sample complexity multiplied by a $O(\ln d)$ factor, i.e.
\begin{align}
    \MC_{\F} = O\left(\frac{dr}{\delta}\ln\frac{d}{\delta}\ln d\right), \quad \MC_{\T} = O\left(\frac{dr}{\epsilon^2}\ln d\right). 
\end{align}
Note the above complexity for the na\"ive application does not take account the overhead of the auxiliary qubits from the continuous POVM, while the complexity for our algorithm does include the overhead of the auxiliary qubits from the discretization.

\subsection{Single-copy strategies}

Single-copy quantum state tomography algorithms have also been actively studied~\cite{Haffner_2005,DAriano_2007,Gross_2010}. 
These algorithms can be classified as non-adaptive and adaptive. In non-adaptive algorithms, measurements do not depend on the outcomes of previous measurements, while in adaptive algorithms, they do. In particular, non-adaptive algorithms can be viewed as special cases of adaptive algorithms and as such, lower bounds for adaptive algorithms are valid for non-adaptive ones as well. 

In this section, we focus on the non-adaptive projection least-square tomography algorithm of~\cite{DAriano_2007}, which is of special interest because it was shown analytically in~\cite[Corollary~1]{Guta_2020} to achieve
\begin{align}
    \SC_{\T} =O\left(\frac{dr^2}{\epsilon^2}\ln d\right),
\end{align}
which is larger by a factor of $r$ than the upper bound of~\cite{ODonnell_2016,Haah_2017}. 

Single-copy tomography algorithms for the infidelity are, however, understudied. 
Only recently in~\cite[Theorem~\Romannum{6}.1]{Chen_2023}, the authors proposed an adaptive single-copy tomography algorithm whose sample complexity is
\begin{align}
    \SC_{\F}=O\left(\frac{dr^2}{\delta}\ln\frac{r}{\delta}\right),
\end{align}
and therefore larger by a factor of $r$ than the fully coherent algorithm of~\cite{ODonnell_2016,Haah_2017}.\footnote{Note that the original proof of~\cite[Theorem~\Romannum{6}.1, Lemma~\Romannum{6}.6]{Chen_2023} states that $\SC_{\F}\big(\delta'\ln \frac{r}{\delta'},d,r\big)=O\big(\frac{dr^2}{\delta'}\big)$, which is equivalent to our statement by properly scaling $\delta = \delta' \ln\frac{r}{\delta'}$, as $\frac{1}{\delta'}\leq \frac{2}{\delta}\ln\frac{r}{\delta}$ asymptotically.}

With respect to lower bounds, it was shown in~\cite[Theorem~\Romannum{3}.11]{Chen_2023} that any adaptive single-copy tomography algorithm with $r=d$ must satisfy
\begin{align}
    \SC_{\T}=\Omega\left(\frac{d^3}{\epsilon^2}\right), 
\end{align}
which subsequently gives
\begin{align}
    \SC_{\F} = \Omega\left(\frac{d^3}{\delta}\right), 
\end{align}
due to the Fuchs-van de Graaf inequality.

It is remarkable that the non-adaptive projection least square algorithm in~\cite{Guta_2020} achieves the $r=d$ lower bound $\SC_{\F} = \Omega\big(\frac{d^3}{\epsilon^2}\big)$ for adaptive single-copy strategies up to a logarithmic factor, revealing that adaptiveness does not yield any improvement in the sample complexity when learning in trace distance. 
However, adaptiveness is indeed necessary for the adaptive single-copy strategy in~\cite[Theorem~\Romannum{3}.11]{Chen_2023} to achieve the $r=d$ lower bound $\SC_{\F} = \Omega\big(\frac{d^3}{\delta}\big)$, which is better than the $r=d$ lower bound $\SC_{\F} = \Omega\big(\frac{d^3}{\delta^2}\big)$ for non-adaptive single-copy strategies in~\cite[Theorem~4]{Haah_2017}. 

Turning to the memory complexity, all single-copy algorithms only store one sample and perform single copy measurements on one qudit, so recalling the $O(\ln d)$ factor when converting from qudits to qubits, we obtain
\begin{align}
    \MC_{\F} = \Theta(\ln d), \quad \MC_{\T} = \Theta(\ln d). 
\end{align}

\subsection{$t$-coherent strategies}
\label{sec:tcoh}
Little was known about quantum state tomography using $t$-coherent algorithms until the recent work~\cite{Chen_2024a}, which presents an algorithm achieving
\begin{align}
    \SC_{\T}=O\left(\frac{d^3}{\sqrt{t} \epsilon^2}(\ln t)^3\right),
\end{align}
independently of the rank $r$ of the input state.\footnote{We remark that the logarithmic factor $O\big((\ln t)^3\big)$ was omitted in the original paper~\cite{Chen_2024a}, but we restore it by going through the proof: \cite[Algorithm 2]{Chen_2024a} calls \cite[Corollary 5.6]{Chen_2024a} $O(\ln t)$ times in order to estimate $\rho$ with $\epsilon$ trace distance. In the $i$-th call of \cite[Corollary 5.6]{Chen_2024a}, the subroutine calls \cite[Algorithm 1]{Chen_2024a} once to estimate the projection of $\rho$ on the subspace of the $i$-th largest eigenvalue within $O\big(\frac{\sqrt{t}}{2^{i-1}}\cdot\frac{\epsilon}{\sqrt{d} \ln t}\big)$ in Frobenius norm, and \cite[Algorithm 1]{Chen_2024a} consumes $O\big(\frac{d^3}{\sqrt{t}\epsilon^2}(\ln t)^2\big)$ samples to achieve such accuracy.} 
However, \cite{Chen_2024a} does not make any claims on the sample complexity with respect to fidelity, and it is unknown whether the same techniques can be used when learning in infidelity.

With regards to lower bounds,~\cite[Theorem 7.1]{Chen_2024a} states that when $r=d$
\begin{align}
    \SC_{\T} = \Omega\left(\frac{d^3}{\sqrt{t} \epsilon^2}\right),
\end{align}
and thus translates to 
\begin{align}
    \SC_{\F} = \Omega\left(\frac{d^3}{\sqrt{t} \delta}\right)
\end{align}
by the Fuchs-van de Graaf inequality. Therefore, the $t$-coherent tomography algorithm of~\cite{Chen_2024a} achieves the $r=d$ lower bound up to a logarithmic factor $O\big((\ln d)^3\big)$. 

In spite of the fact that the $t$-coherent algorithm of~\cite{Chen_2024a} is optimal up to logarithmic factors when $r=d$, it is unknown whether it is optimal when $r\neq d$. This is because the algorithm in~\cite{Chen_2024a} converts the input state to an intermediate state which is close to the fully mixed state, thereby losing any advantage found when learning low rank states. 
Obtaining the upper and lower trade-off bounds between $\SC$ and $t$ when $r\neq d$ remains an open question.  

Concerning the memory complexity: the $t$-coherent algorithm stores $t$ samples and performs $t$-coherent measurements on $t$ qudits, so together with the $O(\ln d)$ factor, we obtain  
\begin{align}
    \MC_{\T} = O(t \ln d). 
\end{align}

As a special case of the $t$-coherent tomography algorithm, we distinguish the $d^2$-coherent tomography algorithm achieving
\begin{align}
    \SC_{\T}= O\left(\frac{d^2}{\epsilon^2}(\ln d)^3\right), 
\end{align}
which is only logarithmically worse than the $r=d$ lower bound $\SC_{\T} = \Omega(\frac{d^2}{\epsilon^2})$ of~\cite{ODonnell_2016}. 
In this case, the memory complexity is 
\begin{align}
    \MC_{\T} = O\big(d^2\ln d\big). 
\end{align}

\subsection{Our setup and advantage}

In repeat-until-success computing schemes~\cite{Lim_2005,Shah_2013}, quantum states are prepared and sent for further processing in a sequential way. Since tomography algorithms receive input states in this manner, it is natural to consider their fit into a streaming algorithmic framework. 
However, in spite of the fruitful research on the sample complexity of quantum state tomography, it is unknown whether the tomography algorithms discussed in previous subsections admit a streaming implementation with lower memory complexity. Indeed, the straight-forward implementation of the fully coherent tomography algorithm in~\cite{ODonnell_2016,ODonnell_2017,Haah_2017} requires $\MC_{\F}=\Omega\big(\frac{dr}{\delta} \ln d \ln \frac{d}{\delta}\big)$.
In contrast, we have $\MC_{\F}=\Omega(\ln d)$-qubit (respectively, $\MC_{\F}=\Omega(t\ln d)$) is required by the single-copy (resp.,$t$-coherent) tomography algorithms of~\cite{Guta_2020,Chen_2023}(resp.,\cite{Chen_2024a}), at the cost of a larger sample complexity. Similar statements hold for $\MC_{\T}$. It is unknown whether memory complexity of these algorithms can be further optimized. 

Our work is a step towards filling this gap by exploring the minimal memory complexity for algorithms with optimal sample complexity. 
We propose a streaming implementation of the fully-coherent tomography algorithms of~\cite{ODonnell_2016,ODonnell_2017,Haah_2017} which achieve the optimal sample complexities in Eq.~\eqref{eq:our-samp-opt} and memory complexities in Eq.~\eqref{eq:our-mem-efficient}.
We compare our algorithm with the fully coherent tomography algorithms in~\cite{ODonnell_2016,ODonnell_2017,Haah_2017} in Table~\ref{tab:comparison_fullycoherent}.
To summarise, our algorithm retains the same sample complexities and obtains an exponential improvement on the memory complexity with respect to the desired accuracies $\delta$ in infidelity and $\epsilon$ in the trace distance, whilst remaining the same in dimension $d$ and rank $r$. 

Another family of tomography algorithm we compare is the family of $t$-coherent tomography algorithms in~\cite{Chen_2024a}. We especially focus on $t$-coherent tomography algorithms with $t = d^2/\polylog(d)$ in Table~\ref{tab:comparison_tcoherent}, as we are searching for minimal memory complexity conditioned on optimal sample complexity, and this family of algorithms saturate the sample complexity lower bound with respect to the trace distance up to poly-logarithmic factors. 

Our algorithm has an advantage over this family in two aspects. When learning in infidelity, our algorithm saturates the sample complexity lower bound up to a logarithmic factor, while the authors of~\cite{Chen_2024a} neither claim any sample complexity results for infidelity nor clarify how their techniques can be applied in this scenario. 
When learning in trace distance, our algorithm performs better than the $t$-coherent in both sample complexity and memory complexity, especially for large dimension $d$ but comparably small rank $r$.
This is because the sample and memory complexity of our algorithm scales as $O\big(\frac{dr}{\epsilon^2}\big)$ and $O\big(dr\polylog(d,\epsilon)\big)$, respectively, whilst those of~\cite{Chen_2024a} scale as $O\big(\frac{d^2}{\epsilon^2}\polylog(d)\big)$ and $O\big(d^2\polylog(d)\big)$, respectively.

\section{Preliminaries}
\label{sec:preliminaries}
\subsection{Distance measures}
For a linear operator $L$ acting on a Hilbert space $\mathcal{H}$, we define the \emph{Schatten $p$-norm} for $p\geq 1$ as
\begin{align}
    \norm{L}_p:= \Big(\Tr|L|^p\Big)^{\frac{1}{p}}.
\end{align}
In fact, $\norm{L}_p$ equals the $\ell_p$ norm of the vector of singular values of $L$.
Of particular interest in this work are the cases $p=1$ and $p=2$, whence the respective Schatten norms are respectively referred to as the \emph{trace norm} and \emph{Frobenius norm} (or \emph{Hilbert-Schmidt norm}).
Additionally, if $\rho, \sigma$ are \emph{quantum states} (positive semidefinite and unit-trace operators on $\mathcal{H}$) then the \emph{trace distance} between $\rho$ and $\sigma$ is 
\begin{align}
    \label{eq:def_trace_distance}
    T(\rho,\sigma):=\frac{1}{2}\norm{\rho-\sigma}_1,
\end{align}
the \emph{fidelity} is
\begin{align}
    F(\rho,\sigma):= \Big(\Tr|\sqrt{\rho}\sqrt{\sigma}|\Big)^2,
\end{align}
and the \emph{infidelity} is 
\begin{align}
    \label{eq:def_infidelity}
    I(\rho,\sigma):= 1 - F(\rho,\sigma).
\end{align}
These distance measures are related by the Fuchs-van de Graaf inequalities
\begin{align}
    1-\sqrt{F(\rho,\sigma)}\leq T(\rho,\sigma)\leq\sqrt{1-F(\rho,\sigma)}.
\end{align}
It is worth noting that the following alternative definitions for the fidelity and the infidelity are also widely used:
\begin{align}
    \tilde{F}(\rho,\sigma) := \Tr|\sqrt{\rho}\sqrt{\sigma}|, \quad \tilde{I}(\rho,\sigma) := 1 - \tilde{F}(\rho,\sigma).  
\end{align}
The two definitions of the infidelity are actually equivalent in the sense that, for any $\rho$ and $\sigma$, 
\begin{align}
    \tilde{I}(\rho,\sigma)\leq I(\rho,\sigma)\leq 2\tilde{I}(\rho,\sigma). 
\end{align}
It is worth noting that the infidelity $I(\rho,\sigma)$ and $\tilde{I}(\rho,\sigma)$ are also related to the purified distance and the Bures distance $B(\rho,\sigma)$ via
\begin{align}
    P(\rho,\sigma) := \sqrt{I(\rho,\sigma)}, \quad B(\rho,\sigma) := \sqrt{2\tilde{I}(\rho,\sigma)}. 
\end{align}
Both Bures distance and purified distance are metrics, i.e., satisfy positivity, symmetry and the triangle inequality. 
In quantum state tomography, the accuracy measure $\Delta(\rho,\hat{\rho})$ is usually chosen to be either the trace distance $T(\rho,\hat{\rho})$ or the infidelity $I(\rho,\hat{\rho})$.

\subsection{Young diagrams and Young tableaux}
A \emph{Young label} is a tuple $\lambda=(\lambda_1,...,\lambda_d)$ satisfying $\lambda_1 \geq ...\geq \lambda_d\geq 0$ and $\sum_{i=1}^d \lambda_i=n$, and is denoted by $\lambda\vdash n$. With a slight abuse of notation, we also use $\lambda$ to denote the diagonal matrix $\diag(\lambda_1,...,\lambda_d)$, and  $\overline{\lambda}$ for the diagonal matrix $\diag(\frac{\lambda_1}{n},...,\frac{\lambda_d}{n})$. 

A Young label $\lambda$ can be associated with a \emph{Young diagram}. For $\lambda\vdash n$, the corresponding Young diagram is an arrangement of $n$ empty boxes into $d$ rows, with $\lambda_i$ boxes in the $i$th row. For a Young diagram $\lambda\vdash n$, a \emph{Young tableau} is an assignment of the integers from $1$ to $n$ into the empty boxes of the Young diagram. A \emph{standard Young tableau} is a Young tableau with $n$ distinct entries strictly increasing along each row and each column. A \emph{semi-standard Young tableau} is a Young tableau with possibly identical entries non-decreasing along each row and strictly increasing along each column. 

\subsection{Schur-Weyl duality}
Schur-Weyl duality is an essential theorem in representation theory which establishes a relation between the irreducible representations (or irreps) of the symmetric group $S(n)$ and those of the general linear group $GL(d)$ (the unitary group $U(d)$ as well)~\cite{Goodman_2000}. 
It states that
\begin{align}\label{eqn:sw_duality}
    (\mathbbm{C}^d)^{\otimes n} \cong \bigoplus_{\lambda\vdash n} \mathcal{P}_\lambda \otimes \mathcal{Q}_\lambda^d, 
\end{align}
where $\lambda\vdash n$ is the Young label, and $\mathcal{P}_\lambda$ and $\mathcal{Q}_\lambda^d$ are respectively the minimal invariant subspaces, i.e., irreducible representations (irrep), of the symmetric group $S(n)$ and the general linear group $GL(d)$. 
The Schur transform $U^n_{\rm Sch}$ on $(\mathbb{C}^d)^{\otimes n}$ is precisely the unitary map underpinning the isomorphism in Eq.~\eqref{eqn:sw_duality}. The basis of $\mathcal{P}_\lambda$ is indexed by standard Young tableaux of Young diagrams of $\lambda$ with at most $d$ rows while the basis of $\mathcal{Q}_\lambda$ is indexed by semi-standard Young tableaux of Young diagrams of $\lambda$. A more detailed introduction on Schur-Weyl duality and its applications to quantum information can be found in~\cite{Harrow_2005,Buhrman_2023,Nguyen_2023,Grinko_2023}.

Due to the permutation invariance of $X^{\otimes n}$ for a given $X\in GL(d)$, one can easily conclude by Schur's lemma~\cite{Goodman_2000}
\begin{align}
    U_{\rm Sch}^n X^{\otimes n} U_{\rm Sch}^{n,\dagger} = \bigoplus_{\lambda \vdash n} \mathbbm{I}_{ \mathcal{P}_\lambda} \otimes {\bm q}_\lambda(X),
\end{align}
where ${\bm q}_\lambda: GL(d)\to GL(\mathcal{Q}_\lambda^d)$ is the irrep of $GL(d)$ indexed by $\lambda$. 
Of particular interest are the projections $\Pi_\lambda$ onto the isotypic subspace indexed by $\lambda$ in the standard basis.
These have the immediate properties that for all $X$ on $(\mathbb{C}^d)^{\otimes n}$
\begin{align}
    U_{\rm Sch}^n\Pi_\lambda^n U_{\rm Sch}^{n,\dagger} & = \mathbbm{I}_{\mathcal{P}_\lambda} \otimes \mathbbm{I}_{\mathcal{Q}_\lambda^d},  \\
    U_{\rm Sch}^n \Pi_\lambda^n X^{\otimes n}\Pi_\lambda^n U_{\rm Sch}^{n,\dagger} & =\mathbbm{I}_{\mathcal{P}_\lambda}\otimes {\bm q}_\lambda(X),\\
    \Pi_\lambda^n X^{\otimes n} & = X^{\otimes n} \Pi_\lambda^n. 
\end{align}

We use $s_\lambda(X)$ to denote the character of the irrep ${\bm q}_\lambda(X)$, that is, for all $X\in GL(d)$:
\begin{align}
    s_\lambda(X)=\Tr( {\bm q}_\lambda(X)).
\end{align}

Lastly, we will state the following facts which are essential for our memory complexity.

\begin{lemma}\label{lem:dimension_bound}
    Let $\mathcal{L}_n^r =\{\lambda\vdash n\,:\,\rank \lambda\leq r\}$ be the set of Young labels with at most $r$ non-zero elements. Then
    \begin{align}
        |\mathcal{L}_n^r| \leq (n+1)^{r}. 
    \end{align} 
    Moreover, for any $\lambda\in\mathcal{L}_n^r$, we have
    \begin{align}
        \dim\mathcal{Q}_\lambda^d \leq (n+1)^{dr}. 
    \end{align}
\end{lemma}

\begin{proof}
    The proof is in Appendix~\ref{apd:combinatorics}. 
\end{proof}

\begin{lemma}\label{lem:majorization}
    Let $U\in SU(d)$. Then we have
    \begin{align}
        \Tr({\bm q}_\lambda(U\overline{\lambda} U^\dagger\rho))>0 \quad \iff \quad \rank\lambda \leq \rank\rho .
    \end{align}
\end{lemma}

\begin{proof}
    Follows from~\cite[Lemma~2]{Haah_2017}, see Appendix~\ref{apd:combinatorics}. 
\end{proof}

\section{POVM construction}
\label{sec:povm_construction}

The fully coherent tomography algorithm in~\cite{ODonnell_2016,Haah_2017} uses a continuous Haar random POVM, which is the main obstacle to properly address its memory and time complexity. Our algorithm is a streaming implementation of the fully coherent tomography algorithm, and thus also has to deal with this obstacle. Therefore, we start by constructing discrete POVMs which emulate the continuous Haar random POVMs and which can also be implemented space efficiently.

\begin{definition}
    \label{def:discrete_povm_equiv}
    Let $\lambda\vdash n$, $\eta\geq 0$ and $\mathcal{S}\subset U(d)$ be finite. Then for $U\in\mathcal{S}$, we define
    \begin{align}
        N_{\mathcal{S},\eta}(\lambda,U):= \frac{\dim \mathcal{Q}_\lambda^d}{(1+\eta)|\mathcal{S}|s_\lambda(\overline{\lambda})} {\bm q}_\lambda(U \overline{\lambda} U^\dagger),
    \end{align}
    and 
    \begin{align}
        N_{\mathcal{S},\eta}(\lambda,{\rm fail}) := \mathbbm{I}_{\mathcal{Q}_\lambda^d} -  \frac{\dim \mathcal{Q}_\lambda^d}{(1+\eta)|\mathcal{S}|s_\lambda(\overline{\lambda})} \sum_{U\in \mathcal{S}}{\bm q}_\lambda(U \overline{\lambda} U^\dagger).
    \end{align}
\end{definition}

It is obvious that 
\begin{align}
    \sum_{U\in\mathcal{S}} N_{\mathcal{S},\eta}(\lambda,U) + N_{\mathcal{S},\eta}(\lambda,{\rm fail}) = \mathbbm{I}_{\mathcal{Q}_\lambda^d}.
\end{align}
In addition, the following lemma shows that $N_{\mathcal{S},\eta}(\lambda,U)$ for $U\in\mathcal{S}$ is always positive.

\begin{lemma}
    \label{lem:irreps_positive}
    Let $\lambda\vdash n$ and let $A\succeq 0$ be a positive operator on $\mathbbm{C}^{d}$. Then ${\bm q}_\lambda(A)\succeq 0$.
\end{lemma}

\begin{proof}
    Since ${\bm q}_\lambda(\cdot)$ is a representation of $GL(d)$, we have
    \begin{align}
        {\bm q}_\lambda(A)={\bm q}_\lambda(\sqrt{A}){\bm q}_\lambda(\sqrt{A}^{\dagger})={\bm q}_\lambda(\sqrt{A}){\bm q}_\lambda(\sqrt{A})^{\dagger}.
    \end{align}
    The last equality is due to the fact that ${\bm q}_\lambda(.)$ commutes with taking the adjoint. Altogether, we can conclude that ${\bm q}_\lambda(A)\succeq 0$.
\end{proof}

In order to ensure $\{N_{\mathcal{S},\eta}(\lambda,U)\}_{U\in\mathcal{S}\cup\{{\rm fail}\}}$ is a POVM, it remains to check whether $N_{\mathcal{S},\eta}(\lambda,{\rm fail}) \succeq 0$ for all $\lambda$. We show that this is indeed the case when $\lambda$, $\eta$ and $\mathcal{S}$ satisfy a certain property. This is the technically difficult part of our proof. 

Before introducing these properties, for partitions $\lambda,\mu$ and irrep ${\bm q}_\mu:GL(d)\to GL(\mathcal{Q}^d_\mu)$ we introduce a type of `twirl' of the operator $U\bar{\lambda}U^\dagger$ within irrep $\mathcal{Q}^d_\mu$. 

\begin{definition}
    Let $\lambda$ and $\mu$ be partitions. Let $\mathcal{S}\subseteq U(d)$ be finite. We define 
    \begin{align}
        \mathcal{U}_{\lambda,\mu} & :=  \int \d U {\bm q}_\mu(U \overline{\lambda} U^\dagger ), \\
        \mathcal{U}_{\lambda,\mu}^{\mathcal{S}} & := \frac{1}{|\mathcal{S}|} \sum_{U \in \mathcal{S}} {\bm q}_\mu(U \overline{\lambda} U^\dagger ).
    \end{align}
    Furthermore $\mathcal{U}_\lambda :=\mathcal{U}_{\lambda,\lambda}$ and $\mathcal{U}_\lambda^{\mathcal{S}} :=\mathcal{U}_{\lambda,\lambda}^{\mathcal{S}} $.
\end{definition}

Armed with this definition, we introduce the following two properties of the set $\mathcal{S}$.

\begin{definition}\label{def:property_S}
    Let $\lambda\vdash n$ be a partition with at most $r$ elements, and let $\mathcal{S} \subseteq U(d)$ be finite. We write $\mathcal{S}\in \mathcal{A}_{n,d}^{r}(\eta,\lambda)$ for some $\eta>0$ if it holds that
    \begin{align}
        (1-\eta) \mathcal{U}_\lambda \preceq \mathcal{U}_\lambda^{\mathcal{S}} \preceq (1+\eta) \mathcal{U}_\lambda,
    \end{align}
    Furthermore we define
    \begin{align}
        \mathcal{A}_{n,d}^{r}(\eta) := \bigcap_{\lambda\vdash n} \mathcal{A}_{n,d}^{r}(\eta,\lambda). 
    \end{align}
\end{definition}

\begin{definition}\label{def:property_T}
    Let $\lambda\vdash n$ be a partition with at most $r$ non-zero elements, $\mu\in\lambda+\Box \vdash (n+1)$ be a partition with at most $r$ non-zero elements obtained by adding a box to $\lambda$. Let $\mathcal{S} \subseteq U(d)$ be finite. Then we denote $\mathcal{S}\in\mathcal{B}_{n,d}^{r}(\eta,\lambda,\mu)$ for some $\eta>0$  if it holds that 
    \begin{align}
        (1-\eta) \mathcal{U}_{\lambda,\mu} \preceq \mathcal{U}_{\lambda,\mu}^{\mathcal{S}} \preceq (1+\eta) \mathcal{U}_{\lambda,\mu},
    \end{align}
    Furthermore we define
    \begin{align}
        \mathcal{B}_{n,d}^{r}(\eta) = \bigcap_{\substack{\lambda\vdash n\\ \mu := \lambda+\Box} }\mathcal{B}_{n,d}^{r}(\eta,\lambda,\mu). 
    \end{align}
\end{definition}

Intuitively, the properties $\mathcal{S}\in \mathcal{A}_{n,d}^{r}(\eta,\lambda)$ and $\mathcal{S}\in\mathcal{B}_{n,d}^{r}(\eta,\lambda,\mu)$ tell us that for the matrix ${\bm q}_\mu(\overline{\lambda})$, the set $\mathcal{S}\subset U(d)$ acts almost like a 1-design on the irrep $\mathcal{Q}_\mu^d$.

Next, we show that there exist a finite set $\mathcal{S}\subset U(d)$ satisfying either Definition \ref{def:property_S} or Definition \ref{def:property_T}.
To do this, we make use of the following operator-valued Chernoff inequality from~\cite[Theorem~19]{Ahlswede_2002}. 
\begin{lemma}
    \label{lem:operator_chernoff}
    Let $\mathcal{H}$ be a Hilbert space and let $X,X_1,...,X_n$ be i.i.d. random Hermitian matrices on $\mathcal{H}$ with values in $[0,\mathbbm{I}]$. Suppose $\expect[X] \geq \zeta \mathbbm{I}$ for some $\zeta>0$ and $\eta\in[0,\frac{1}{2}]$, then
    \begin{align}
        \Pr\left[\frac{1}{n}\sum_i X_i \notin \big[(1-\eta)\expect[X], (1+\eta) \expect[X]\big]\right] \nonumber \\ 
        \leq 2\dim\mathcal{H}  \exp\left(-\frac{\zeta \eta^2 n}{2\ln 2}\right).
    \end{align}
\end{lemma}

Using this lemma, we propose an explicit algorithm to construct $\mathcal{S}$ such that $\mathcal{S}\in\mathcal{A}_{n,d}^{r}(\eta)$ or $\mathcal{S}\in\mathcal{B}_{n,d}^{r}(\eta)$ with $|\mathcal{S}|=O\big(\frac{dr}{\eta^2}n^{dr}\ln n\big)$ elements. 
\begin{theorem}\label{thm:povm_lower_bound}
    Let $\eta\in[0,\frac{1}{2}]$ and let $\mathcal{S}\subseteq U(d)$ be a finite set whose elements are i.i.d. according to the Haar measure on $U(d)$. Let $n\geq d\geq2$ and $r\geq 1$. Lastly, let $\lambda\vdash n$ and $\mu\in\lambda+\ydiagram{1}$. Then, the following two properties hold: 
    \begin{enumerate}
        \item[(1)] It holds that 
    \begin{align}
        & \Pr[\mathcal{S}\in \mathcal{A}_{n,d}^{r}(\eta)]\nonumber\\
        & \geq 1-2(n+1)^{2dr}\exp{\left(-\frac{\eta^2|\mathcal{S}| }{2 (n+1)^{dr} \ln2}\right)}.
    \end{align}
    In particular, $\Pr[\mathcal{S}\in \mathcal{A}_{n,d}^{r}(\eta)]>0$ provided 
    \begin{align}
        |\mathcal{S}|= \left\lceil\frac{2(n+1)^{dr}\ln2}{\eta^2}\ln\big(2(n+1)^{2dr}\big)\right\rceil,
    \end{align}
    which means $|\mathcal{S}|=O\big(\frac{dr n^{dr}}{\eta^2}\ln n\big)$. 
    \item[(2)] It holds that 
    \begin{align}
        & \Pr[\mathcal{S}\in \mathcal{B}_{n,d}^{r}(\eta)] \nonumber\\
        & \geq 1-2(n+2)^{2dr}\exp{\left(-\frac{\eta^2|\mathcal{S}|}{2 (n+2)^{dr}\ln2}\right)}.
    \end{align}
    In particular, $\Pr[\mathcal{S}\in \mathcal{B}_{n,d}^{r}(\eta)]>0$ provided 
    \begin{align}
        |\mathcal{S}|=\left\lceil\frac{2(n+2)^{dr}\ln2}{\eta^2}\ln\big(2(n+2)^{2dr}\big)\right\rceil,
    \end{align}
    which means $|\mathcal{S}|=O\big(\frac{dr n^{dr}}{\eta^2}\ln n\big)$. 
    \end{enumerate}    
\end{theorem}

\begin{proof}
    To show bounds on the probabilities that $\mathcal{S}\in\mathcal{A}_{n,d}^{r}(\eta)$ or that $\mathcal{S}\in\mathcal{B}_{n,d}^{r}(\eta)$, we want to eventually apply the operator-valued Chernoff bound from Lemma~\ref{lem:operator_chernoff} to operators of the form ${\bm q}_\nu(U\overline{\lambda}U^\dagger)$, where $\nu$ is a placeholder for either $\lambda$ or $\mu$ as given in the theorem statement. However, Lemma~\ref{lem:operator_chernoff} requires that $\expect[{\bm q}_\nu(U\overline{\lambda}U^\dagger)]>0$.
    For starters, note 
    \begin{align}
        \Tr{\expect[{\bm q}_\nu(U\overline{\lambda}U^\dagger)]} = \expect[\Tr({\bm q}_\nu(U\overline{\lambda}U^\dagger))] = s_\nu(\overline{\lambda}),
    \end{align}
    so combining with the positivity of $\expect[{\bm q}_\nu(U\overline{\lambda}U^\dagger)]$ indicated by Lemma~\ref{lem:irreps_positive}, we have that $\expect[{\bm q}_\nu(U\overline{\lambda}U^\dagger)]>0$ if and only if $s_\nu(\overline{\lambda})>0$. 
    Hence, we need to consider the case where $s_\nu(\overline{\lambda})=0$ and the case where $s_\nu(\overline{\lambda})>0$. 
    
    Firstly, if $s_\nu(\overline{\lambda})=0$, then 
    \begin{align}
        \Tr(\mathcal{U}_{\lambda,\nu}) = \Tr(\mathcal{U}_{\lambda,\nu}^{\mathcal{S}}) = s_\nu(\overline{\lambda}) = 0,
    \end{align}
    so the positivity of $\mathcal{U}_{\lambda,\nu}$ and $\mathcal{U}_{\lambda,\nu}^{\mathcal{S}}$ implies
    \begin{align}
        \mathcal{U}_{\lambda,\nu} = \mathcal{U}_{\lambda,\nu}^{\mathcal{S}} = 0. 
    \end{align}
    Hence it trivially follows that $\mathcal{S}\in \mathcal{A}_{n,d}^r(\eta,\lambda)$ and $\mathcal{S}\in \mathcal{B}_{n,d}^r(\eta,\lambda,\mu)$. 
    Thus, in the remainder of the proof it suffices to assume $s_\nu(\overline{\lambda})\neq 0$ to consider the cases when $\nu=\lambda$ (Statement (1)) and $\nu=\mu$ (Statement (2)). 
    
    We begin with Statement (1). Let $U_i\in\mathcal{S}$ and define 
    \begin{align}\label{eq:x_i}
        X^i & :=\frac{{\bm q}_\lambda(U_i \overline{\lambda} U_i^\dagger )}{s_{\lambda}(\overline{\lambda})}, \\
        X & := \frac{1}{|\mathcal{S}|}\sum_{U_i\in\mathcal{S}} X^i.\label{eq:x}
    \end{align}
    Lemma \ref{lem:irreps_positive} tells us that ${\bm q}_\lambda(\overline{\lambda})\succeq 0$, so we have $X^i \succeq 0$. On the other hand, we recall that $s_\lambda(X) = \Tr({\bm q}_\lambda(X))$, and we notice that 
    \begin{align}
        \Tr{\mathbbm{E}[X^i]} = \frac{1}{s_\lambda(\overline{\lambda})}\mathbbm{E}\left[\Tr({\bm q}_\lambda(U_i\overline{\lambda}U_i^\dagger))\right]=1 .
    \end{align}
    Here, the expectation is with respect to $U_i$ sampled according to the Haar measure. Together with $X^i \succeq 0$ this implies $X^i\in[0,\mathbbm{I}_{\mathcal{Q}_\lambda^d}]$.
    
    Further, due to the unitary invariance of the Haar measure, we have that for any unitary $V\in U(d)$
    \begin{align}
        \mathbbm{E}[X^i] ={\bm q}_\lambda(V)\mathbbm{E}[X^i]{\bm q}_\lambda(V^\dagger). 
    \end{align}
    Together with $\Tr{\mathbbm{E}[X^i]}=1$, Schur's Lemma implies
    \begin{align}
        \mathbbm{E}[X]= \mathbbm{E}[X^i]=\frac{\mathbbm{I}_{\mathcal{Q}_\lambda^d}}{\dim\mathcal{Q}_{\lambda}^d} . 
    \end{align}
    As $\mathcal{S}\in \mathcal{A}_{n,d}^r(\eta,\lambda)$ if and only if
    \begin{align}
        (1-\eta)\expect[X] \preceq X \preceq (1+\eta)\expect[X], 
    \end{align}
    we can use Lemma \ref{lem:operator_chernoff} with $\zeta=\frac{1}{\dim\mathcal{Q}_{\lambda}^d}$ to obtain
    \begin{align}
        \Pr\left[\mathcal{S}\notin \mathcal{A}_{n,d}^{r}(\eta,\lambda) \right] \leq 2\dim\mathcal{Q}_{\lambda}^d  \exp\left(-\frac{\eta^2|\mathcal{S}|}{2\ln 2\dim\mathcal{Q}_{\lambda}^d}\right).
    \end{align}
    Since we have restricted $\lambda$ to have at most $r$ non-zero elements as is required by the property $\mathcal{A}_{n,d}^{r}(\eta)$ and $\mathcal{B}_{n,d}^{r}(\eta)$, we obtain $\dim\mathcal{Q}_{\lambda}^d\leq(n+1)^{dr}$ which we note is independent of $\lambda$.
    Finally, by the union bound, 
    \begin{align}
        & \Pr[\mathcal{S}\notin \mathcal{A}_{n,d}^r(\eta)] \leq\sum_{\lambda\vdash n}\Pr\left[\mathcal{S}\notin \mathcal{A}_{n,d}^r(\eta,\lambda)\right] \nonumber  \\
        & \leq \sum_{\lambda\vdash n}2(n+1)^{dr} \exp\left(-\frac{\eta^2|\mathcal{S}|}{2\ln 2(n+1)^{dr}}\right),
    \end{align}
    where in the last inequality we noted that $|\{\lambda\in\mathbbm{N}_+^r:\lambda\vdash n\}|\leq (n+1)^{r}$. This concludes the proof of (1).
    
    We now turn to Statement (2). Let $U_i\in\mathcal{S}$ and for $\mu$ as in the statement of the theorem let
    \begin{align}
        X^i & :=\frac{{\bm q}_\mu(U_i \overline{\lambda} U_i^\dagger )}{s_{\mu}(\overline{\lambda})}, \\
        X & := \frac{1}{|\mathcal{S}|}\sum_{U_i\in\mathcal{S}} X^i.
    \end{align}
    Proceeding as before, $\Tr{\expect[X^i]}=1$ implies that $ X^i\in [0, \mathbbm{I}_{\mathcal{Q}_\mu^d}]$, and the unitary invariance of the integration over the Haar measure together with $\Tr{\expect[X^i]}=1$, we again obtain
    \begin{align}
        \expect[X] = \expect[X^i] = \frac{\mathbbm{I}_{\mathcal{Q}_\mu^d}}{\dim \mathcal{Q}_\mu^d}.
    \end{align}
    Similarly, $\mathcal{S}\in \mathcal{B}_{n,d}^r(\eta,\lambda,\mu)$ if and only if 
    \begin{align}
        (1-\eta)\expect[X]\preceq X \preceq (1+\eta)\expect[X], 
    \end{align}
    so Lemma \ref{lem:operator_chernoff} with $\zeta=\frac{1}{\dim\mathcal{Q}_{\mu}^d}$ yields
    \begin{align}
        \Pr[\mathcal{S}\notin\mathcal{B}_{n,d}^r(\eta,\lambda,\mu)]\leq 2\dim\mathcal{Q}_{\mu}^d  \exp\left(-\frac{\eta^2|\mathcal{S}|}{2\ln 2\dim\mathcal{Q}_{\mu}^d}\right). 
    \end{align}
    Using $\dim \mathcal{Q}_\mu^d \leq (n+2)^{dr}$, we obtain
    \begin{align}
        & \Pr[\mathcal{S}\notin \mathcal{B}_{n,d}^r(\eta)] \leq\sum_{\substack{\lambda\vdash n\\ \mu\in\lambda + \Box}}\Pr\left[\mathcal{S}\notin \mathcal{B}_{n,d}^r(\eta,\lambda,\mu)\right] \nonumber  \\
        & \leq \sum_{\substack{\lambda\vdash n\\ \mu\in\lambda+\Box}}2(n+2)^{dr} \exp\left(-\frac{\eta^2|\mathcal{S}|}{2\ln 2(n+2)^{dr}}\right),
    \end{align}
    where again, we noted that $|\{\lambda\in\mathbbm{N}_+^r:\lambda\vdash n\}|\leq (n+1)^{r}$ as well as $|\{\mu:\mu\in\lambda+\Box\}|\leq r$. This concludes the proof of (2). 
\end{proof}

Finally, we prove that $\{N_{\mathcal{S},\eta}(\lambda,U)\}_{U\in\mathcal{S}\cup\{{\rm fail}\}}$ where either $\mathcal{S}\in\mathcal{A}_{n,d}^{r}(\eta)$ or $\mathcal{S}\in\mathcal{B}_{n,d}^{r}(\eta)$ is indeed a POVM.
\begin{lemma}
    \label{lem:discrete_povm}
    Let $\lambda\vdash n$, $\mathcal{S}\subseteq U(d)$ be finite and $\eta\in(0,1)$ be so that $\mathcal{S}\in \mathcal{A}_{n,d}^r(\eta)$ or $\mathcal{S}\in \mathcal{B}_{n,d}^r(\eta)$. Then $\{N_{\mathcal{S},\eta}(\lambda,U)\}_{U\in\mathcal{S}\cup\{{\rm fail}\}}$ in Definition~\ref{def:discrete_povm_equiv} is a POVM in both cases. 
\end{lemma}

\begin{proof}
    (1) Consider the case where $\mathcal{S}\in \mathcal{A}_{n,d}^r(\eta)$. 
    First, notice that from the unitary invariance of $\mathcal{U}_\lambda$ we have $[\mathcal{U}_\lambda, {\bm q}_\lambda(V)]=0$ for all $\lambda\vdash n$ and $V\in U(d)$, such that Schur's lemma implies $\mathcal{U}_\lambda=\frac{s_\lambda(\overline{\lambda})}{\dim\mathcal{Q}_{\lambda}^d}\mathbbm{I}_{\mathcal{Q}_\lambda^d}$. 
    Further, since $\mathcal{S}\in \mathcal{A}_{n,d}^r(\eta)$, we have
    \begin{align}
        \frac{(1-\eta)\mathbbm{I}_{\mathcal{Q}_\lambda^d} }{\dim\mathcal{Q}_{\lambda}^d}
        \preceq \frac{\mathcal{U}_\lambda^{\mathcal{S}}}{s_{\lambda}(\overline{\lambda})} \preceq \frac{(1+\eta)\mathbbm{I}_{\mathcal{Q}_\lambda^d}}{\dim\mathcal{Q}_{\lambda}^d}\,,
    \end{align}
    so multiplying by $\dim\mathcal{Q}_{\lambda}^d$, dividing by $(1+\eta)$ and using the definition of $\mathcal{U}_\lambda^{\mathcal{S}}$ gives 
    \begin{align}
        \label{eqn:discrete_povm_element_bound}
        \frac{1-\eta}{1+\eta}\mathbbm{I}_{\mathcal{Q}_\lambda^d} 
        \preceq \frac{\dim\mathcal{Q}_{\lambda}^d}{(1+\eta)|\mathcal{S}|}  \sum_{U \in \mathcal{S}} \frac{{\bm q}_\lambda(U \overline{\lambda} U^\dagger )}{s_{\lambda}(\overline{\lambda})}
        \preceq \mathbbm{I}_{\mathcal{Q}_\lambda^d}.
    \end{align}
    Since $\eta\in(0,1)$, we can deduce that $N_{\mathcal{S},\eta}(\lambda,{\rm fail})$ is a positive operator, and therefore that $\{N_{\mathcal{S},\eta}(\lambda,U)\}_{U\in\mathcal{S}\cup\{{\rm fail}\}}$ is a POVM. 
    
    (2) Consider now the case where $\mathcal{S}\in \mathcal{B}_{n,d}^r(\eta)$. 
    Using an auxiliary system, let
    \begin{align}\label{eq:wlambda}
        \mathcal{W}_{\lambda} & = \int \d U {\bm q}_\lambda(U\overline{\lambda}U^\dagger)\otimes U\overline{\lambda}U^\dagger, \\
        \mathcal{W}_{\lambda}^{\mathcal{S}} & = \frac{1}{|\mathcal{S}|} \sum_{U_i\in\mathcal{S}} {\bm q}_\lambda(U\overline{\lambda}U^\dagger)\otimes U\overline{\lambda}U^\dagger. \label{eq:wslambda}
    \end{align}
    The Clebsch-Gordan transform~\cite{Bacon_2006,Bacon_2007} implies
    \begin{align}
        \label{eqn:CG_transform}
        {\bm q}_\lambda(U\overline{\lambda}U^\dagger)\otimes U\overline{\lambda}U^\dagger \simeq \bigoplus_{\mu = \lambda+\Box} {\bm q}_\mu(U\overline{\lambda}U^\dagger),
    \end{align}
    hence, noting that ${\bm q}_\mu(U \overline{\lambda}U^\dagger)=0$ when $\mu$ has more than $r$ non-zero elements~\cite[Lemma~2]{Haah_2017}:
    \begin{align}
        \mathcal{W}_\lambda & \simeq \bigoplus_{\mu\in\lambda+\Box} \int \d U {\bm q}_\mu(U\overline{\lambda}U^\dagger) = \bigoplus_{\mu\in\lambda+\Box} \mathcal{U}_{\lambda,\mu}, \\
        \mathcal{W}_\lambda^{\mathcal{S}} & \simeq \bigoplus_{\mu\in\lambda+\Box} \frac{1}{|\mathcal{S}|} \sum_{U\in\mathcal{S}}{\bm q}_\mu(U\overline{\lambda}U^\dagger) = \bigoplus_{\mu\in\lambda+\Box} \mathcal{U}_{\lambda,\mu}^{\mathcal{S}}.
    \end{align}
    Now, since by assumption $\mathcal{S}\in \mathcal{B}_{n,d}^r(\eta)$, it holds that for all $\lambda\vdash n$ and $\mu\in\lambda+\Box$
    \begin{align}
        (1-\eta) \mathcal{U}_{\lambda,\mu}
        \preceq \mathcal{U}_{\lambda,\mu}^{\mathcal{S}} \preceq (1+\eta) \mathcal{U}_{\lambda,\mu}.
    \end{align}
    After summing over $\mu\in\lambda+\Box$, it holds that 
    \begin{align}
        \label{eqn:CG_decompose}
        (1-\eta) \mathcal{W}_{\lambda}
        \preceq \mathcal{W}_{\lambda}^{\mathcal{S}} \preceq (1+\eta) \mathcal{W}_{\lambda}.
    \end{align}
    Lastly, taking the partial trace over the auxiliary system in the last inequality, $\mathcal{W}_\lambda$ and $\mathcal{W}_\lambda^{\mathcal{S}}$ reduce to $\mathcal{U}_{\lambda}$ and $\mathcal{U}_\lambda^{\mathcal{S}}$ respectively, so we conclude that
    \begin{align}
        (1-\eta) \mathcal{U}_{\lambda}
        \preceq \mathcal{U}_{\lambda}^{\mathcal{S}} \preceq (1+\eta) \mathcal{U}_{\lambda},
    \end{align}
    for all $\lambda\vdash n$.
    It follows that $\mathcal{S}\in \mathcal{A}_{n,d}^{r}(\eta)$ and that $\{N_{\mathcal{S},\eta}(\lambda,U)\}_{U\in\mathcal{S}\cup\{{\rm fail}\}}$ is a POVM, concluding the proof. 
\end{proof}

We remark that the POVM $\{N_{\mathcal{S},\eta}(\lambda,U)\}_{U\in\mathcal{S}\cup\{{\rm fail}\}}$ measures on $\mathcal{Q}_\lambda^d$ and as such is not suitable to analyze the overall performance of our later tomography algorithm.
For this reason we introduce the following POVM $\{M_{\mathcal{S},\eta}(\lambda,U)\}_{U\in\mathcal{S}\cup\{{\rm fail}\}}$ on $U_{\rm Sch}^{n,\dagger} (\mathcal{P}_\lambda \otimes \mathcal{Q}_\lambda^d) U_{\rm Sch}^{n} $.

\begin{definition}
    \label{def:discrete_povm}
    Let $\eta\geq 0$ and $\mathcal{S}\subset U(d)$ be finite. Then for $\lambda\vdash n$ and $U\in\mathcal{S}$, we define
    \begin{align}
        M_{\mathcal{S},\eta}(\lambda,U) 
        & :=\frac{\dim \mathcal{Q}_\lambda^d}{(1+\eta)|\mathcal{S}|s_\lambda(\overline{\lambda})} \Pi_\lambda^n (U\overline{\lambda} U^\dagger)^{\otimes n} \Pi_\lambda^n, \\
        M_{\mathcal{S},\eta}(\lambda,{\rm fail}) 
        & := \Pi_\lambda^n - \sum_{U\in\mathcal{S}} M_{\mathcal{S},\eta}(\lambda,U), 
    \end{align}
    in the standard basis and thus 
    \begin{align}
        U_{\rm Sch}^n M_{\mathcal{S},\eta}(\lambda,U) U_{\rm Sch}^{n,\dagger} & = \mathbbm{I}_{\mathcal{P}_\lambda} \otimes N_{\mathcal{S},\eta}(\lambda,U), \\
        U_{\rm Sch}^n M_{\mathcal{S},\eta}(\lambda,{\rm fail}) U_{\rm Sch}^{n,\dagger} & =  \mathbbm{I}_{\mathcal{P}_\lambda} \otimes N_{\mathcal{S},\eta}(\lambda,{\rm fail}), 
    \end{align}
    in the Schur basis. 
\end{definition}

Again, it is obvious that 
\begin{align}
    \sum_{\lambda\vdash n} \left(\sum_{U\in\mathcal{S}} M_{\mathcal{S},\eta}(\lambda,U) + M_{\mathcal{S},\eta}(\lambda,{\rm fail})\right) = \mathbbm{I}_{d^n}. 
\end{align}

We remark that an identical proof to Lemma~\ref{lem:discrete_povm} can be used to show that if $\mathcal{S}\in\mathcal{A}^r_{n,d}(\eta)$ or $\mathcal{S}\in\mathcal{B}^r_{n,d}(\eta)$, the set $\{M_{\mathcal{S},\eta}(\lambda,U)\}_{\lambda\vdash n, U\in\mathcal{S}\cup\{{\rm fail}\}}$ is a POVM as well.
Indeed, the difficulty also lies in showing that $M_{\mathcal{S},\eta}(\lambda,{\rm fail})$ is a positive operator.

We highlight that generating the set $\mathcal{S}$ will not negatively impact the quantum memory complexity of our protocol. Because $\mathcal{S}$ is a finite, fixed set, it can be sampled entirely offline before the protocol begins. The i.i.d. sampling analyzed in our work is simply a probabilistic method to guarantee the existence of a finite set $\mathcal{S}$ that satisfies the the desired properties $\mathcal{A}_{n,d}^r(\eta)$ or $\mathcal{B}_{n,d}^r(\eta)$. We will not rely on algorithmic randomness during the execution of the tomography protocol itself. The properties of $\mathcal{S}$ are deterministic once found. It only needs to be sampled from the Haar measure and verified its properties once ever, permanently stored in classical memory, and simply looked up whenever the algorithm is run. Therefore, any computational or memory overhead associated with forming $\mathcal{S}$ is strictly classical and pre-computational, leaving the operational quantum memory complexity of the streaming phase unaffected.

\section{A streaming algorithm for quantum state tomography}
\label{sec:algorithm}
One of the main ingredients in our memory efficient and sample optimal tomography algorithm is the memory efficient protocol for unitary Schur sampling from~\cite{Cervero_2024}, presented in Algorithm~\ref{alg:stream_wss}. Unitary Schur sampling is defined in~\cite{Cervero_2024} as the task equivalent to first measuring the irrep label $\lambda$ while keeping the post-measurement state on $\mathcal{P}_{\lambda} \otimes \mathcal{Q}_{\lambda}^{d}$, and then discarding the permutation register $\mathcal{P}_\lambda$.

In the algorithm, for each $\lambda\vdash k$ we use $\lambda+\ydiagram{1}:=\{\mu\vdash {k+1}\,:\,\mu=\lambda+\mathbf{e}_j \textrm{ for } j\in[d]\}$ to denote the set of valid Young labels with one additional box.

\begin{figure}[h]
\begin{algorithm}[H]\label{alg:stream_wss}
\SetAlgoLined
\caption{A streaming and memory efficient algorithm for unitary Schur sampling}
\SetKwInOut{Input}{Input}
\SetKwInOut{Output}{Output}
\SetKwInOut{Init}{Initialization}
\SetKwRepeat{Repeat}{repeat}{until}
\Input{A stream of an $n$-qudit state $\sigma$ on $\mathbbm{C}^{d^n}$ \\
($Q_k$ is the $k$-th qudit register) }
\Output{Partition $\lambda^n\vdash n$\\
Post-measurement state $\sigma'$ on $\mathcal{Q}_{\lambda^n}^d$}
Initialization\;
$R_1 = Q_1$\;
$\lambda^1 = (1)$\;
\For{$k=1$ to $n-1$}{
Receive the $(k+1)$-th qudit $Q_{k+1}$\;
Apply Clebsch-Gordan transform $U_{\rm CG}^{\lambda^k}$ on $R_k\otimes Q_{k+1}$\;
Measure to obtain the partition $\lambda^{k+1}\vdash(k+1)$ satisfying $\lambda^{k+1}\in\lambda^k+\ydiagram{1}$\;
Set $R_{k+1}$ to be the post-measurement register whose Hilbert space is $\mathcal{Q}_{\lambda^{k+1}}^d$.
}
Set $\sigma'$ to be the state over register $R_n$ in space $\mathcal{Q}_{\lambda^n}^{d}$\;
\Return $\lambda^n$, $\sigma'$
\end{algorithm}
\end{figure}

In particular it is shown in~\cite{Cervero_2024} that the unitary Schur sampling protocol in Algorithm~\ref{alg:stream_wss} requires only a working memory of $d^2\ln n$ qubits (or $dr\ln n$ qubits if the input is $\sigma=\rho^{\otimes n}$ with $\rank(\rho)=r$), an exponential improvement (in $n$) over naïve algorithms for unitary Schur sampling using the Schur transform.

\begin{lemma}[\cite{Cervero_2023}, Remark 1]
    \label{lem:stream_wss}
    Let $\sigma= \rho^{\otimes n}$ be an $n$-qudit state. Consider the probability distribution $p$ over labels $\lambda\vdash d$ with $p_{\lambda} = \Tr[\sigma \Pi^n_\lambda]$.
    Then Algorithm~\ref{alg:stream_wss} with input $\sigma$ outputs a partition $\lambda\vdash d$ and a state $\tilde{\sigma}_\lambda$ with a probability $\tilde{p}_{\lambda}$ such that $T(\tilde{p}, p) \leq \xi$ and $T\big(\frac{1}{s_\lambda(\rho)}{\bm q}_\lambda(\rho), \tilde{\sigma}_\lambda\big)\leq \xi$. This algorithm has a memory complexity of $O(dr\ln n)$ and a time complexity of $O(d^3r n^{2dr+1}\polylog(d,n,\frac{1}{\xi}))$.
\end{lemma}


A more precise but technical version of Lemma~\ref{lem:stream_wss} is found in Lemma~\ref{lem:stream_wss_restated} in the appendix.

We remark that we use the modified version given in Remark 1 of~\cite{Cervero_2024} rather than Theorem 1. The former algorithm has a logarithmically better memory complexity but an exponentially worse time complexity compared to the latter. However, as the POVM measurement in our tomography algorithm has even worse time efficiency, this difference in time complexity is irrelevant. If this measurement is made time-efficient in the future, then we can instead apply the main algorithm in~\cite{Cervero_2024}.

At this point, we make the subtle observation that a call to Algorithm \ref{alg:stream_wss} followed by an application of POVM $\{N_{\mathcal{S},\eta}(\lambda,U)\}_{U\in\mathcal{S}\cup\{{\rm fail}\}}$ to obtain respective measurement outcomes $\lambda$ and $U$ is equivalent to a direct application of POVM $\{M_{\mathcal{S},\eta}(\lambda,U)\}_{\lambda\vdash n, U\in\mathcal{S}\cup\{{\rm fail}\}}$ which directly produces outcomes $(\lambda, U)$.

\begin{lemma}
    \label{lem:povm_equiv}
    For input $\rho^{\otimes n}$, the probability of obtaining $(\lambda,U)$ by either measuring with POVM $\{M_{\mathcal{S},\eta}(\lambda,U)\}_{\lambda\vdash n, U\in\mathcal{S}\cup\{{\rm fail}\}}$, or first measuring $\lambda$ with Algorithm~\ref{alg:stream_wss} followed by measuring $U$ with $\{N_{\mathcal{S},\eta}(\lambda,U)\}_{U\in\mathcal{S}\cup\{{\rm fail}\}}$, are the same. 
    That is,
    \begin{align}
        \Pr_{N_{\mathcal{S},\eta}}[\lambda,U] = \Pr_{M_{\mathcal{S},\eta}}[\lambda,U].
    \end{align}
\end{lemma}
\begin{proof}
    By Lemma~\ref{lem:stream_wss}, the probability of obtaining $(\lambda,U)$ by measuring $\lambda$ with Algorithm~\ref{alg:stream_wss} followed by measuring $U$ with $N_{\mathcal{S},\eta}$ is 
    \begin{align}
        & \Pr_{N_{\mathcal{S},\eta}}[\lambda,U]  = \Pr_{N_{\mathcal{S},\eta}}[\lambda] \Pr_{N_{\mathcal{S},\eta}}[U|\lambda] \nonumber \\
        & = \dim (\mathcal{P}_\lambda) s_\lambda(\rho) \Tr[s_\lambda(\rho)^{-1} {\bm q}_\lambda(\rho) N_{\mathcal{S},\eta}(\lambda,U) ] \nonumber \\
        & = \Tr[(\mathbbm{I}_{\mathcal{P}_\lambda} \otimes {\bm q}_\lambda(\rho))(\mathbbm{I}_{\mathcal{P}_\lambda}\otimes N_{\mathcal{S},\eta}(\lambda,U))] \nonumber \\
        & = \Tr[\Pi_{\lambda}^n\rho^{\otimes n}\Pi_{\lambda}^n M_{\mathcal{S},\eta}(\lambda,U)] \nonumber \\
        & = \Tr[\rho^{\otimes n} M_{\mathcal{S},\eta}(\lambda,U)] = \Pr_{M_{\mathcal{S},\eta}}[\lambda,U], 
    \end{align}
    which is exactly the probability of obtaining $(\lambda,U)$ by measuring $(\lambda,U)$ with $M_{\mathcal{S},\eta}(\lambda,U)$. 
\end{proof}

This lemma allows us to analyse an alternative but equivalent version of Algorithm \ref{alg:stream_tomo} which forgoes the execution of the unitary Schur sampling algorithm, Algorithm \ref{alg:stream_wss}, and applies the POVM $\{M_{\mathcal{S},\eta}(\lambda,U)\}_{\lambda\vdash n, U\in\mathcal{S}\cup\{{\rm fail}\}}$ directly.

Finally, this allows us to introduce our sample-optimal, memory-efficient tomography protocol in Algorithm \ref{alg:stream_tomo}.

\begin{figure}[h]
\begin{algorithm}[H]\label{alg:stream_tomo}
\SetAlgoLined
\caption{A memory efficient tomography algorithm}
\SetKwInOut{Input}{Input}
\SetKwInOut{Output}{Output}
\SetKwInOut{Init}{Initialization}
\SetKwRepeat{Repeat}{repeat}{until}
\Input{$n$ qudit state $\rho^{\otimes n}$\\
Finite subset $\mathcal{S}\subset U(d)$\\
Parameter $\eta\geq 0$}
\Output{Classical description $\hat{\rho}$ of $\rho$}
Initialization\;
Apply Algorithm \ref{alg:stream_wss} to $\sigma=\rho^{\otimes n}$, to obtain $\lambda\vdash n$ and post-measurement state $\sigma'=\frac{1}{s_\lambda(\rho)}{\bm q}_\lambda(\rho)$\;
Apply POVM $\{N_{\mathcal{S},\eta}(\lambda, U)\}_{U\in\mathcal{S}}\cup\{N_{\mathcal{S},\eta}(\lambda, {\rm fail})\}$ to $\frac{1}{s_\lambda(\rho)}{\bm q}_\lambda(\rho)$, to obtain $U\in \mathcal{S}\cup\{{\rm fail}\}$\;
\eIf{$U={\rm fail}$}{
Output $\hat{\rho}(\lambda,{\rm fail})=\frac{1}{d}\mathbbm{I}_d$.\;
}{
Output $\hat{\rho}(\lambda,U)=U\bar{\lambda}U^\dagger$.
}
\end{algorithm}
\end{figure}
We make a couple of remarks regarding this algorithm.
Firstly, we note that sampling of the finite subset $\mathcal{S}$ is likely efficient due to recent developments in constructions of random unitaries with low-depth circuits \cite{Ma_2025, Schuster_2025}.
Secondly, note that by construction in Section \ref{sec:povm_construction}, the set $\mathcal{S}$ is independent of the unknown state $\rho$. Consequently, the corresponding POVM $\big\{N_{\mathcal{S},\eta}(\lambda, U)\big\}_{U\in\mathcal{S}}\cup\big\{N_{\mathcal{S},\eta}(\lambda, {\rm fail})\big\}$ may be `standardised' and reused in independent executions of Algorithm \ref{alg:stream_tomo} for different unknown states. 
In particular, this allows us to omit the complexity of sampling $\mathcal{S}$ from the complexity analysis of Algorithm \ref{alg:stream_tomo} since it only needs to be performed once. See Appendix~\ref{apd:reconstruction} for more details. 

We now investigate the sample and memory complexity of Algorithm \ref{alg:stream_tomo} with respect to infidelity and trace distance.

\subsection{Complexity with respect to infidelity}\label{sec:infidelity}

In this section, we prove upper bounds on the sample, memory and time complexity of our streaming quantum state tomography algorithm when learning in infidelity.
We begin by proving an upper bound on the infidelity between the input state $\rho$ and the output $\hat{\rho}$.

\begin{lemma}
    \label{lem:discrete_povm_infidelity}
    Let $\eta\in(0,1)$, $\mathcal{S}\in\mathcal{A}_{n,d}^{r}(\eta)$ be a finite set and $\{M_{\mathcal{S},\eta}(\lambda,U)\}_{\lambda\vdash n, U\in\mathcal{S}\cup\{{\rm fail}\}}$ be the POVM in Definition~\ref{def:discrete_povm}. Further, let $\rho$ be a state on $\mathbbm{C}^d$ with $\rank(\rho)\leq r$ and let $\hat{\rho}(\lambda,U)$ be the state assigned to the measurement outcome $U$ of $\{M_{\mathcal{S},\eta}(\lambda,U)\}_{\lambda\vdash n, U\in\mathcal{S}\cup\{{\rm fail}\}}$, that is
    \begin{align}
        \hat{\rho}(\lambda,U) :=U\overline{\lambda}U^{\dagger},\quad
        \hat{\rho}(\lambda,{\rm fail}) :=\frac{1}{d}\mathbbm{I}_d.
    \end{align}
    Then we have for $\delta\geq 0$ that
    \begin{align}
        \label{eq:infidelity_inequality_general}
        \Pr[I(\hat{\rho},\rho) \leq \frac{\delta}{2}]\geq 1- 2\eta -(n+1)^{3dr} \left(1-\frac{\delta}{2}\right)^{n},
    \end{align}
    Further, assuming $n\geq 2d$ and setting $n= O\big(\frac{dr}{\delta}\ln \frac{d}{\delta} + \frac{1}{\delta}\ln \frac{1}{\xi}\big)$ and $\eta =O(\xi)$, we obtain
    
    \begin{align}
        \label{eq:infidelity_inequality_sample_complexity}
        \Pr[I(\hat{\rho},\rho)\leq \delta] \geq 1-\xi, 
    \end{align}
    which means 
    \begin{align}
        \SC_{\F} = O\left(\frac{dr}{\delta} \ln \frac{d}{\delta} +\frac{1}{\delta} \ln\frac{1}{\xi}\right). 
    \end{align}
\end{lemma}

\begin{proof}
    Since $\mathcal{S}\in\mathcal{A}_{n,d}^{r}(\eta)$, we can use a similar argument as in Eq.~\eqref{eqn:discrete_povm_element_bound} to get
    \begin{align}
        \frac{1-\eta}{1+\eta}\Pi_\lambda^n\preceq\sum_{U\in \mathcal{S}} M_{\mathcal{S},\eta}(\lambda,U).
    \end{align}
    This gives us
    \begin{align}
        M_{\mathcal{S},\eta}(\lambda,{\rm fail}) 
        \preceq \frac{2\eta}{1+\eta} \Pi_\lambda^n.
    \end{align}
    Since we know that
    \begin{align}
        \sum_{\lambda\vdash n} \Pi_\lambda^n=\id_{d^n},
    \end{align}
    we find that
    \begin{align}
        \Pr[{\rm fail}]\leq\frac{2\eta}{1+\eta}\quad{\rm and}\quad\Pr[{\rm success}]\geq \frac{1-\eta}{1+\eta}. 
    \end{align} 
    Now, from~\cite[Eq.~(12)]{Haah_2017} (see~\footnote{In~\cite{Haah_2017}, the fidelity between $\rho$ and $\sigma$ are defined as the square-root fidelity $\tilde{F}(\rho,\sigma)$} for explanation) it is known that 
    \begin{align}
        & \Tr\big[M_{\mathcal{S},\eta}(\lambda, U)\rho^{\otimes n}\big] \nonumber\\
        & \leq \frac{1}{(1+\eta)|\mathcal{S}|} (n+1)^{2dr} F(\rho,\hat{\rho}(\lambda,U))^{n},
    \end{align}
    where $r$ is the rank of $\rho$. We can sum over all outcomes $\lambda\vdash n$ and $U\in\mathcal{S}$ to obtain
    \begin{align}
        & \Pr[I(\hat{\rho},\rho)\geq \frac{\delta}{2}] \nonumber \\
        &\leq \sum_{\substack{\lambda\vdash n \\ U\in\mathcal{S}}}\frac{1}{(1+\eta)|\mathcal{S}|}(n+1)^{2dr} \left(1-\frac{\delta}{2}\right)^{n} + \Pr[{\rm fail}] \nonumber \\
        &\leq \frac{1}{1+\eta}(n+1)^{3dr} \left(1-\frac{\delta}{2}\right)^{n} + \frac{2\eta}{1+\eta}\,,
    \end{align} 
    where in the last line we used $|\{\lambda:\lambda\vdash n\}|\leq(n+1)^d$. 
    This concludes the proof for Equation \eqref{eq:infidelity_inequality_general}. 

    To find the sample complexity, we set $n=\frac{12dr}{\delta}\ln \frac{6dr}{\delta} + \frac{2}{\delta}\ln \frac{2}{\xi}$ and $\eta = \frac{\xi}{4}$ to obtain 
    \begin{align}
        \Pr[I(\hat{\rho},\rho)\leq \delta ] \geq 1 - \xi,  
    \end{align}
    which concludes the proof of the lemma. 
\end{proof}

Using the sample complexity $\SC_{\F}$ in Lemma~\ref{lem:discrete_povm_infidelity}, we derive the memory complexity $\MC_{\F}$ of Algorithm~\ref{alg:stream_tomo}. 

\begin{theorem} \label{thm:tomo_memo_sample_F}
    Algorithm \ref{alg:stream_tomo} has the following sample and memory complexities with respect to infidelity:
    \begin{align}
        \SC_{\F}=O\left(\frac{dr}{\delta}\ln \frac{d}{\delta}\right), \quad \MC_{\F}=O\left(dr\ln\frac{d}{\delta}\right). 
    \end{align}
\end{theorem}

\begin{proof}
    We use Lemma~\ref{lem:discrete_povm_infidelity}, note that $\xi$ is constant and get $n=O\big(\frac{dr}{\delta}\ln \frac{dr}{\delta}\big)$ as well as 
    \begin{align}
        \SC_{\F}=O\left(\frac{dr}{\delta}\ln\frac{d}{\delta}\right). 
    \end{align}
    By Lemma~\ref{lem:povm_equiv}, we can implement Algorithm~\ref{alg:stream_tomo} by first measuring $\lambda$ with Algorithm~\ref{alg:stream_wss} and then measuring $U$ with the POVM $N_{\mathcal{S},\eta}$. We now analyze the two parts of our algorithm respectively. 
    
    By Lemma~\ref{lem:stream_wss}, Algorithm~\ref{alg:stream_wss} requires $O(dr\ln n)$ qubits. Substituting $n$, the memory complexity is
    \begin{align}
        \MC_{\F}^{\rm Sch} =O\left(dr\ln\frac{d}{\delta}\right), 
    \end{align}
    for the unitary Schur sampling. 
    
    By Theorem~\ref{thm:povm_lower_bound}, we can find $\mathcal{S}$ with $|\mathcal{S}|=O\big(\frac{drn^{dr}}{\eta^2}\ln n\big)$. As is shown in Lemma~\ref{lemma:approximate_implementation_measurement} and~\ref{lem:povm_unitary}, the POVM $N_{\mathcal{S},\eta}$ can then be implemented with $O\big(\ln (\frac{drn^{dr}}{\eta^2}\ln n)\big)$ auxiliary qubits, while only changing $\xi$ to $3\xi$.

    We know from Lemma~\ref{lem:discrete_povm_infidelity} that $\eta=O(\xi)$. Substituting $n$ and $\eta$, the contribution to the memory complexity for the POVM is
    \begin{align}
        \MC_{\F}^{\rm M}=O\left(dr\ln\frac{d}{\delta}\right). 
    \end{align}
    
    Taking the maximum between $\MC_{\F}^{\rm Sch}$ and $\MC_{\F}^{\rm M}$ gives the overall memory complexity
    \begin{align}
        \MC_{\F}=O\left(dr\ln\frac{d}{\delta}\right),
    \end{align}
    as required. 
\end{proof}

For the sake of completeness, we also compute the time complexity of our algorithm.

\begin{lemma}
    The time complexity of Algorithm \ref{alg:stream_tomo} is
    \begin{align}
        \TC_{\F}  =O\left((dr)^3\left(\frac{dr }{\delta}\right)^{4dr} \polylog\left(d,\frac{1}{\delta}\right)\right).
    \end{align}
    Furthermore, the time complexity is dominated by the POVM. 
\end{lemma}
\begin{proof}
    We set $n=O(\frac{dr}{\delta}\ln\frac{d}{\delta})$ and $\eta=O(\xi)$.
    
    By Lemma~\ref{lem:stream_wss}, the number of elementary one- and two-qubit gates that Algorithm~\ref{alg:stream_wss} needs is 
    \begin{align}
        \TC^{\rm Sch} = O\left(d^3r n^{2dr}\polylog\left(d,n,\frac{1}{\xi}\right)\right). 
    \end{align}
    Substituting $n$ into this expression implies that the time complexity for the unitary Schur sampling algorithm is 
    \begin{align}
        \TC_{\F}^{\rm Sch} = O\left(d^3 r\left(\frac{dr}{\delta}\right)^{2dr}\polylog\left(d,\frac{1}{\delta}\right)\right). 
    \end{align}

    By Lemma~\ref{lemma:approximate_implementation_measurement} and~\ref{lem:povm_unitary}, the POVM $N_{\mathcal{S},\eta}$ can be implemented with 
    \begin{align}
        \TC^{\rm M} =O\left( (\dim\mathcal{Q}_\lambda^d |\mathcal{S}|)^2 \ln \frac{\dim\mathcal{Q}_\lambda^d |\mathcal{S}|}{\xi}\right), 
    \end{align}
    elementary gates. On the one hand, $\dim\mathcal{Q}_\lambda^d \leq (n+1)^{dr}$ by Lemma~\ref{lem:dimension_bound}. On the other hand, $|\mathcal{S}|=O(\frac{drn^{dr}}{\eta^2}\ln n)$ by Theorem~\ref{thm:povm_lower_bound}. So we can substitute $n$ and $\eta$ into $\dim\mathcal{Q}_\lambda^d$ and $|\mathcal{S}|$ to get a gate complexity of 
    \begin{align}
        \TC_{\F}^{\rm M} = O\left((dr)^3\left(\frac{dr }{\delta}\right)^{4dr} \left(\ln \frac{d}{\delta}\right)^{4dr+3}\right), 
    \end{align}
    for the discrete POVM measurement.

    In conclusion, the time complexity of the full algorithm is obtained by summing $\TC_{\F}^{\rm Sch}$ and $\TC_{\F}^{\rm M}$:
    \begin{align}
        \TC_{\F}  = O\left((dr)^3\left(\frac{dr }{\delta}\right)^{4dr} \left(\ln \frac{d}{\delta}\right)^{4dr+3}\right),
    \end{align}
    which is clearly dominated by the discrete POVM. 
\end{proof}

\subsection{Complexity with respect to trace norm}\label{sec:trace_norm}
This section follows a similar plan as Section~\ref{sec:infidelity}, except that all theorems and lemmas are proven for the trace distance rather than the infidelity. We first prove Lemma~\ref{lem:discrete_povm_trace}, which sets an upper bound on the second moment of the trace distance between $\hat{\rho}$ and $\rho$.

\begin{lemma}
    \label{lem:discrete_povm_trace}
    Let $\eta\in(0,1)$, $\mathcal{S}\in\mathcal{B}_{n,d}^r(\eta)$ be a finite set and $\{M_{\mathcal{S},\eta}(\lambda,U)\}_{\lambda\vdash n, U\in\mathcal{S}\cup\{{\rm fail}\}}$ be the POVM in Definition~\ref{def:discrete_povm}. Further, let $\rho$ be a state on $\mathbbm{C}^d$ with $\rank(\rho)\leq r$ and let $\hat{\rho}(\lambda,U)$ be the state assigned to the measurement outcome $U$ of $\{M_{\mathcal{S},\eta}(\lambda,U)\}_{\lambda\vdash n, U\in\mathcal{S}\cup\{{\rm fail}\}}$, that is
    \begin{align}
        \hat{\rho}(\lambda,U) := U\overline{\lambda}U^{\dagger},\quad \hat{\rho}(\lambda,{\rm fail}) := \frac{1}{d}\mathbbm{I}_d.
    \end{align}
    Then we have 
    \begin{align}
        \expect\big[\|\hat{\rho}-\rho\|_1^2|{\rm success}\big] & \leq \frac{8r(d+\eta n)}{(1-\eta)n},\\
        \Pr[{\rm fail}] & \leq \frac{2\eta}{1+\eta}. 
    \end{align}
\end{lemma}
\begin{proof}
    First of all, as is shown in Lemma~\ref{lem:discrete_povm}, $\mathcal{S}\in \mathcal{B}_{n,d}^r(\eta)$ implies $\mathcal{S}\in \mathcal{A}_{n,d}^r(\eta)$. 
    Hence, Eq.~\eqref{eqn:discrete_povm_element_bound} once again implies
    \begin{align}
        \Pr[{\rm fail}]\leq\frac{2\eta}{1+\eta},\quad\Pr[{\rm success}]\geq \frac{1-\eta}{1+\eta}. 
    \end{align} 
    It then follows that  
    \begin{align}
        & \expect\big[\|\hat{\rho} - \rho\|_2^2 \big| {\rm success}\big] \nonumber \\
        & =\frac{1}{\Pr[{\rm success}]}\sum_{\substack{\lambda\vdash n \\ U\in\mathcal{S}}} \Pr[\lambda,U] \|U\overline{\lambda}U^\dagger - \rho\|_2^2  \\
        & \leq  \frac{1+\eta}{1-\eta}\sum_{\substack{\lambda\vdash n \\ U\in\mathcal{S}}}   \Tr(M_{\mathcal{S},\eta}(\lambda,U)\rho^{\otimes n}) \Tr((U\overline{\lambda}U^\dagger - \rho)^2) \nonumber,  
    \end{align}
    where we have lower-bounded the success probability with $\frac{1-\eta}{1+\eta}$. Continuing the above:
    \begin{align}
        & \expect \big[\|\hat{\rho} - \rho\|_2^2 \big|{\rm success}\big] \leq \sum_{\substack{\lambda\vdash n \\ U\in\mathcal{S}}}  \frac{\dim \mathcal{Q}_\lambda \dim \mathcal{P}_\lambda}{(1-\eta)|\mathcal{S}| s_\lambda(\overline{\lambda})} \nonumber  \\
        & \quad  \cdot\Tr( {\bm q}_\lambda(U\overline{\lambda}U^\dagger \rho )  ) \Tr(\overline{\lambda}^2 + \rho^2 - 2 U \overline{\lambda}U^\dagger \rho ).
    \end{align}
    Noting that 
    \begin{align}
        \Tr( {\bm q}_\lambda (U \overline{\lambda} U^\dagger \rho)  ) \Tr(U \overline{\lambda} U^\dagger \rho) =  \Tr( {\bm q}_\lambda (U \overline{\lambda} U^\dagger \rho)  \otimes U \overline{\lambda} U^\dagger \rho),
    \end{align}
    we can further simplify the expectation value into 
    \begin{align}
        & \expect\big[\|\hat{\rho} - \rho\|_2^2 \big|{\rm success}\big] \leq \sum_{\lambda\vdash n }  \frac{\dim \mathcal{Q}_\lambda \dim \mathcal{P}_\lambda}{(1-\eta) s_\lambda(\overline{\lambda})} \nonumber  \\
        & \cdot \Big(\Tr(\mathcal{U}_\lambda^{\mathcal{S}}{\bm q}_\lambda(\rho))\Tr(\overline{\lambda}^2 + \rho^2) -2\Tr(\mathcal{W}_\lambda^{\mathcal{S}} {\bm q}_\lambda(\rho)\otimes \rho )\Big).
    \end{align}
    Then, using the fact that $\mathcal{S}\in \mathcal{B}_{n,d}^r(\eta)$, we obtain
    \begin{align}
        \Tr(\mathcal{U}_\lambda^{\mathcal{S}}{\bm q}_\lambda(\rho)) \leq  (1+\eta) \Tr(\mathcal{U}_\lambda {\bm q}_\lambda(\rho)),
    \end{align}
    where $\mathcal{W}^\mathcal{S}_\lambda$ is as per Eq.~\eqref{eq:wslambda}.
    Further, recalling that $\mathcal{S}\in \mathcal{B}_{n,d}^r(\eta)$ implies $\mathcal{S}\in \mathcal{A}_{n,d}^r(\eta)$, we use Eq.~\eqref{eqn:CG_decompose} to bound
    \begin{align}
         \Tr( \mathcal{W}_\lambda^{\mathcal{S}} {\bm q}_\lambda(\rho)\otimes \rho) \geq (1-\eta)\Tr( \mathcal{W}_\lambda {\bm q}_\lambda(\rho)\otimes \rho),
    \end{align}
    where $\mathcal{W}_\lambda$ is as in Eq.~\eqref{eq:wlambda}.
    Therefore,
    \begin{align}
        & \expect\big[\|\hat{\rho} - \rho\|_2^2 \big| {\rm success}\big] \leq  \sum_{\lambda\vdash n} \frac{\dim \mathcal{Q}_\lambda \dim \mathcal{P}_\lambda}{(1-\eta)s_\lambda(\overline{\lambda})} \cdot \nonumber \\
        & \myquad[4]\Big((1+\eta)\Tr(\mathcal{U}_\lambda {\bm q}_\lambda (\rho)) \Tr(\overline{\lambda}^2+\rho^2) \nonumber \\
        & \myquad[6]  - (1-\eta)\Tr(\mathcal{W}_\lambda {\bm q}_\lambda (\rho)\otimes \rho)  \Big) \nonumber \\
        & = \frac{1}{1-\eta}(\expect_{\rm ctn}\big[\|\hat{\rho}-\rho\|_2^2\big] + \eta \expect_{\rm ctn}\big[\|\hat{\rho}+\rho\|_2^2\big]),
    \end{align}
    where $\expect_{\rm ctn}$ is the expectation when the measurement is continuous over the Haar measure. 
    Now, it is shown in~\cite{ODonnell_2016,ODonnell_2016b,ODonnell_2017} that 
    \begin{align} 
        \expect_{\rm ctn}\big[\|\hat{\rho}-\rho\|_2^2\big]  \leq \frac{4d-3}{n}, 
    \end{align}
    and since we have
    \begin{align}
        \expect_{\rm ctn}\big[\|\hat{\rho}+\rho\|_2^2\big]  \leq 2 \expect_{\rm ctn}[\Tr(\hat{\rho}^2 + \rho^2)] \leq 4, 
    \end{align}
    it follows that 
    \begin{align}
        \expect\big[\|\hat{\rho}-\rho\|_2^2 \big|{\rm success}\big] \leq \frac{4d-3}{(1-\eta)n} + \frac{4\eta}{1-\eta}.
    \end{align}
    Lastly, from Lemma~\ref{lem:majorization}, $\rank\lambda \leq \rank\rho$, which implies $\rank(\hat{\rho}) \leq r$ and $\rank(\hat{\rho}-\rho)\leq 2r$ such that the Cauchy-Schwarz inequality applied to the singular value decomposition of $\hat{\rho}-\rho$ gives 
    \begin{align}
        \expect\big[\|\hat{\rho} -\rho\|_1^2 \big|{\rm success}\big] \leq 2r \expect\big[\|\hat{\rho} -\rho\|_2^2 \big|{\rm success}\big]. 
    \end{align}
    Combining all these inequalities, we have 
    \begin{align}
         \expect\big[\|\hat{\rho} -\rho\|_1^2 \big|{\rm success}\big] \leq \frac{8rd}{(1-\eta)n} + \frac{8r\eta}{1-\eta},
    \end{align}
    which proves the bound of the expectation value.
\end{proof}

Lemma \ref{lem:discrete_povm_trace} gives us a good estimate for the SMD condition, which is discussed in more detail in Appendix~\ref{apd:cond_equiv}. We now want to translate this condition into the PAC condition we have used so far.

\begin{corollary}
    \label{cor:discrete_povm_trace_complexity}
    In the setting of Lemma \ref{lem:discrete_povm_trace}, assuming $n\geq d$ and setting $n=O\big(\frac{dr}{\epsilon^2}\big)$ and $\eta=O\big(\frac{\epsilon^2}{r}\big)$, we can perform $O(\ln\frac{2}{\xi})$ repetitions of such measurements to obtain 
    \begin{align}
        \Pr[T(\hat{\rho},\rho)\leq \epsilon] \leq 1- \xi, 
    \end{align}
    which means 
    \begin{align}
        \SC_{\T} = O\left(\frac{dr}{\epsilon^2}\ln\frac{2}{\xi}\right). 
    \end{align}
\end{corollary}

\begin{proof}
    Let $\eta\leq \frac{1}{16}$. Using a similar argument as in the proof for Lemma~\ref{lem:pac_repetition}, we can first perform $\frac{2\ln 2/\xi}{\ln 2}$ repetitions of such measurements, and with a probability of at least $1- \frac{1}{2}\xi$, there are $\frac{\ln 2/\xi}{\ln 2}$ successful measurements, that is, 
    \begin{align}
        \Pr[{\rm enough\ success}] \geq 1-\frac{1}{2}\xi.
    \end{align}
    Conditioned on enough successful measurements, using Lemma~\ref{lem:cond_equiv}, we can use $\frac{\ln 2/\xi}{\ln 2}$ repetitions of such successful measurements to obtain 
    \begin{align}
        \Pr[T(\rho,\sigma)\leq 30\sqrt{\frac{r(d+\eta n)}{2(1-\eta)n}}] \geq 1- \frac{1}{2}\xi. 
    \end{align}
    
    To find the sample complexity, we set $n=\frac{900 dr}{\epsilon^2}$ and $\eta = \frac{\epsilon^2}{1200r}$ to obtain
    \begin{align}
        \Pr[T(\rho,\sigma)\leq \epsilon] \geq 1- \xi, 
    \end{align}
    concluding the proof. 
\end{proof}

With the sample complexity $\SC_{\T}$ in Corollary~\ref{cor:discrete_povm_trace_complexity} in hand, we can proceed to the memory complexity $\MC_{\T}$. 

\begin{theorem} \label{thm:tomo_memo_sample_T}
    Algorithm \ref{alg:stream_tomo} has the following sample and memory complexities with respect to trace distance:
    \begin{align}
        \SC_{\T}=O\left(\frac{dr}{\epsilon^2}\right),  \quad \MC_{\T}=O\left(dr\ln\frac{d}{\epsilon}\right).
    \end{align}
\end{theorem}

\begin{proof}
    We proceed as in the proof of Theorem~\ref{thm:tomo_memo_sample_F}. Setting $\xi$ to be constant, we apply Lemma~\ref{lem:discrete_povm_trace} to get $n=O(\frac{dr}{\epsilon^2})$ as well as 
    \begin{align}
        \SC_{\T}=O\left(\frac{dr}{\epsilon^2}\right). 
    \end{align}
    
    Substituting $n$ and $\eta$ into Lemma~\ref{lem:stream_wss}, the memory complexity for the unitary Schur sampling algorithm is
    \begin{align}
        \MC_{\T}^{\rm Sch} =O\left(dr\ln\frac{d}{\epsilon}\right).  
    \end{align}
    
    Combining Theorem~\ref{thm:povm_lower_bound} and Lemma~\ref{lemma:approximate_implementation_measurement} as well as~\ref{lem:povm_unitary}, we can find a $\mathcal{S}$ with $|\mathcal{S}|=O\big(\frac{drn^{dr}}{\eta^2}\ln n\big)$ implementable with $O\big(\ln (\frac{drn^{dr}}{\eta^2}\ln n)\big)$ auxiliary qubits with a failure probability of at most $3\xi$.

    Finally, we know from Lemma~\ref{lem:discrete_povm_trace} that $\eta=O(\frac{\epsilon^2}{r})$ respectively. Substituting $n$ and $\eta$ for the trace distance, the memory complexity for the POVM is
    \begin{align}
        \MC_{\T}^{\rm M} =O\left(dr\ln\frac{d}{\epsilon}\right). 
    \end{align}
    
    Summing $\MC_{\T}^{\rm Sch}$ and $\MC_{\T}^{\rm M}$, we obtain the overall memory complexity
    \begin{align}
        \MC_{\T}=O\left(dr\ln\frac{d}{\epsilon}\right), 
    \end{align}
    which ends our proof.  
\end{proof}

We also mention the time complexity in this case.
\begin{lemma}
    The time complexity of Algorithm~\ref{alg:stream_tomo} is
    \begin{align}
        \TC_{\T} = O\left(\frac{d^3r^7}{\epsilon^8}\left(\frac{dr }{\epsilon^2}\right)^{4dr} \left(\ln \frac{d}{\epsilon}\right)^3\right), 
    \end{align}
    Furthermore, the time complexity is dominated by the POVM. 
\end{lemma}

\begin{proof}
    The proof for the trace distance is similar to that for the infidelity. We set $n=O(\frac{dr}{\epsilon^2})$ and $\eta=O(\frac{\epsilon^2}{r})$ as before.
    Substituting $n$ into Lemma~\ref{lem:stream_wss}, the time complexity of Algorithm~\ref{alg:stream_wss} for the unitary Schur sampling is 
    \begin{align}
        \TC_{\T}^{\rm Sch} = O\left(d^3 r\left(\frac{dr}{\epsilon^2}\right)^{2dr}\polylog\left(d,\frac{1}{\epsilon}\right)\right).
    \end{align}
    
    We can bound $\dim\mathcal{Q}_\lambda^d$ by Lemma~\ref{lem:dimension_bound} and $|\mathcal{S}|$ by Theorem~\ref{thm:povm_lower_bound}. Substituting $n$ and $\eta$ into $\dim\mathcal{Q}_\lambda^d$ and $|\mathcal{S}|$ in Lemma~\ref{lemma:approximate_implementation_measurement} and~\ref{lem:povm_unitary}, we can obtain a gate complexity of 
    \begin{align}
        \TC_{\T}^{\rm M} = O\left(\frac{d^3r^7}{\epsilon^8}\left(\frac{dr }{\epsilon^2}\right)^{4dr} \left(\ln \frac{d}{\epsilon}\right)^3\right), 
    \end{align}
    for the discrete POVM measurement.

    In conclusion, the time complexity of the full algorithm is dominated by the contribution from the discrete POVM measurement part $\TC_{\T}^{\rm M}$, i.e., 
    \begin{align}
        \TC_{\T} = O\left(\frac{d^3r^7}{\epsilon^8}\left(\frac{dr }{\epsilon^2}\right)^{4dr} \left(\ln \frac{d}{\epsilon}\right)^3\right), 
    \end{align}
    as what we state in the lemma. 
\end{proof}

\section{Discussion}
\label{sec:discussion}
Our algorithm, in spite of its sample optimality and memory efficiency, has not been optimised with respect to time efficiency. 
Indeed this remains the pertaining issue for sample-optimal practical quantum state tomography algorithms.
The hardness in optimizing the time complexity of our algorithm lies in the implementation of the `discrete Haar measurement' $\{{N}_{\mathcal{S},\eta}(\lambda,U)\}_{U\in\mathcal{S}\cup\{{\rm fail}\}}$ over the large set $\mathcal{S}\subset U(d)$. We note that there are time-efficient algorithms to apply Haar random unitaries~\cite{Haferkamp_2022,Harrow_2023,Haah_2024, Metger_2024,Schuster_2025}. However, there is a non-trivial gap between applying Haar random unitaries and measuring Haar random POVM elements.

We also remark that because the limiting factor in the time complexity likely lies in the discrete measurement rather than in the unitary Schur sampling protocol, we choose the algorithm in~\cite{Cervero_2023} which has a logarithmically better memory complexity but exponentially worse time complexity than that in~\cite{Cervero_2024}.
If the discrete measurement is made time efficient, the unitary Schur sampling can also be made time efficient by applying the algorithm in~\cite{Cervero_2024} at the cost of a logarithmic increase in the memory complexity. 
We leave the time-efficient implementation of the discrete measurement to future work.

As mentioned in the introduction, we focus only on optimizing the quantum memory of state tomography, since quantum memory is usually much more expensive than classical memory. The optimization of classical memory is another interesting problem, which has been discussed for example in~\cite{Hou_2016}.

There are two frequently used conditions to characterize the accuracy of quantum state tomography algorithms. 
The first, which we also use in this work, is the PAC condition~\cite{Guta_2020,Haah_2017,Chen_2023,Chen_2024a}. The other one, the SMD condition, is found in e.g.~\cite{ODonnell_2016,ODonnell_2017,Flammia_2024,Chen_2024a}. It measures the second moment of the distance measure $\sqrt{\expect[\Delta(\rho,\hat{\rho})^2]}\leq \epsilon$. However, we show in Appendix \ref{apd:cond_equiv} that both conditions are essentially equivalent. As such, all our results could be reframed in the SMD condition by multiplying with a factor of $O(\ln\epsilon)$. On the other hand, any result for the SMD condition translates into a result for the PAC condition by multiplying with a factor $O(\ln\xi)$. 

\begin{acknowledgments}
We thank Josep Lumbreras for useful discussions. 
MT, YH and ECM are supported by the National Research Foundation, Singapore and A*STAR under its CQT Bridging Grant. YH is also funded by the National Research Foundation, Singapore and A*STAR under its Quantum Engineering Programme (NRF2021-QEP2-01-P06) and MT is supported by the NRF Investigatorship award (NRF-NRFI10-2024-0006). ET and LM are supported by European Union via ERC grant (QInteract, Grant No 101078107) and  VILLUM FONDEN (Grant No 10059 and 37532).
\end{acknowledgments}

\bibliography{reference}

\appendix

\section{Some combinatorics}
\label{apd:combinatorics}

We first equip ourselves with the following lemma. 
\begin{lemma}
    \label{lem:dividing}
    The number of ways to divide $n$ boxes into $r$ groups is upper bounded by $(n+1)^{r-1}$. 
\end{lemma}
\begin{proof} 
    Suppose that we put $n$ boxes in a line and then place $r-1$ dividers to the left or the right of any of the $n$ boxes, allowing more than one divider to be placed in the same position. There are $n+1$ positions that we can insert dividers. These $r-1$ dividers divide $n$ boxes into $r$ or less groups depending on if more than one dividers are inserted into the same position. Each divider has $n+1$ possible positions and $r-1$ dividers have no more than $(n+1)^{r-1}$ possible configurations. 
\end{proof}
With the help of the previous lemma, we can proceed to the proof of Lemma~\ref{lem:dimension_bound}. 
\begin{proof}[Proof of Lemma~\ref{lem:dimension_bound}]
    In the language of Young tableaux, $|\mathcal{L}_n^r|$ is upper bound by the number of ways to divide $n$ boxes into $r$ or less groups (without constraining the relative length of the groups), which is upper bounded by $(n+1)^{r}$ as is shown in Lemma~\ref{lem:dividing}.

    It is known \cite{Goodman_2000, Bacon_2006,Bacon_2007} that $\dim\mathcal{Q}_\lambda^d$ equals the number of semi-standard Young tableaux associated with the partition $\lambda\vdash n$ and entries in $[d]$. The number of ways to fill the $i$th row is upper bounded by the number of ways to divide $\lambda_i$ boxes into $d$ or less groups, which is no more than $(\lambda_i+1)^d$ by Lemma~\ref{lem:dividing}. 
    Since there are at most $r$ different rows for elements $\lambda \in \mathcal{L}^r_n$, the number of semi-standard Young tableaux is upper bounded by $(n+1)^{dr}$, as required.
\end{proof}
To prove Lemma~\ref{lem:majorization}, we will first remind the reader of~\cite[Lemma~2]{Haah_2017}. 
\begin{lemma}{\cite[Lemma~2]{Haah_2017}}
    Let $\rho$ and $\sigma$ be density matrices on $\mathbbm{C}^d$. Suppose $\rho$ has rank $r$. Then, the character function $s_\lambda(\rho\sigma)$ satisfies 
    \begin{align}
        s_\lambda(\rho\sigma) \begin{cases}
            \leq \dim \mathcal{Q}_\lambda^de^{-2nH(\overline{\lambda})}F^{n}, & \lambda_{r+1}=0, \\
            =0, & \lambda_{r+1}>0, 
        \end{cases}
    \end{align}
    where
    \begin{align}
        F = F(\rho,\sigma) = \Tr(|\sqrt{\rho}\sqrt{\sigma}|)^2. 
    \end{align}
\end{lemma}
Then Lemma~\ref{lem:majorization} is obvious from~\cite[Lemma~2]{Haah_2017}. 
\begin{proof}[Proof of Lemma~\ref{lem:majorization}]
    It trivially holds from~\cite[Lemma~2]{Haah_2017} because when $\rank\lambda> \rank\rho$, $\Tr({\bm q}_\lambda(U\overline{\lambda}U^\dagger)\rho)=0$. 
\end{proof}

\section{Accuracy conditions for quantum state tomography}
\label{apd:cond_equiv}
For a distance measure $\Delta$, input state $\rho$, output estimate $\hat{\rho}$ and accuracy $\epsilon$, and a failure probability $\xi$, the \emph{probably approximately correct} (PAC) condition~\cite{Guta_2020,Haah_2017,Chen_2023,Chen_2024a} is characterised by
\begin{align}
    \Pr[\Delta(\rho,\hat{\rho})\leq \epsilon]\geq 1-\xi. 
\end{align}
In a similar way, the \emph{second moment of distance} (SMD) condition~\cite{ODonnell_2016,ODonnell_2017,Flammia_2024,Chen_2024a} is given by
\begin{align}
    \sqrt{\expect[\Delta(\rho,\hat{\rho})^2]}\leq \epsilon.
\end{align}
Although we have stated all our results via the PAC condition, we have actually already used the SMD condition in Lemma~\ref{lem:discrete_povm_trace}. In the following we argue that both approaches are equivalent up to logarithmic factors.

We assume that $\Delta$ is a metric in this section. A metric satisfies identity, positivity, symmetry and the triangle inequality, respectively, for any distinct states $\rho$, $\sigma$ and $\tau$ 
\begin{align}
    \Delta(\rho,\rho) & = 0, \\
    \Delta(\rho,\sigma) & > 0, \\
    \Delta(\rho,\sigma) & = \Delta(\sigma,\rho), \\
    \Delta(\rho,\sigma) & \leq \Delta(\rho,\tau) +  \Delta(\sigma,\tau). 
\end{align}
The trace distance is a metric, and Lemma~\ref{lem:pac_repetition} and Lemma~\ref{lem:cond_equiv} hold for the trace distance. Both the purified distance and the Bures distance are metrics, and Lemma~\ref{lem:pac_repetition} and Lemma~\ref{lem:cond_equiv} hold for the purified distance and the Bures distance, and thus the infidelity as well. 

We first show that the failure probability of the PAC condition can be suppressed exponentially with repetitions. It is worth noting that the average of all estimations, bounded with the Bernstein inequality or the Hoeffding bound, concentrates slowly and leads to $O(\frac{1}{\epsilon})$ repetitions due to its small expectation and variance, which forces us to use the median of means method inspired by~\cite{Lumbreras_2024a,Lumbreras_2024b}. 

\begin{lemma}
    \label{lem:pac_repetition}
    If an algorithm can obtain an estimation $\hat{\rho}$ for the state $\rho$ to $\epsilon$ accuracy with a probability of $1-\xi$ where $\xi< \frac{1}{4}$, i.e.
    \begin{align}
        \Pr[\Delta(\rho,\hat{\rho})\leq \epsilon]\geq 1-\xi,
    \end{align}
    then $2c$ repetitions of the algorithm can obtain an estimation $\hat{\rho}$ for the state $\rho$ to $3\epsilon$ accuracy with a probability exponentially close to $1$, i.e.
    \begin{align}
        \Pr[\Delta(\rho,\hat{\rho})\leq 3\epsilon]\geq 1-(4\xi)^c.
    \end{align}
\end{lemma}
\begin{proof}
    We repeat the algorithm for $2c$ times and obtain the estimation $\hat{\rho}_i$ for $i\in[2c]$. We now take $\hat{\rho}_{i^*}$ which maximizes the number of $\rho_j$ inside its $2\epsilon$-ball (that is, $\Delta(\hat{\rho}_{i^*},\hat{\rho}_j)\leq 2\epsilon$). We now show that $\hat{\rho}_{i^*}$ is a good estimation with a probability exponentially close to $1$. 
    
    First, the probability that at least $c+1$ estimations are inside the $\epsilon$ ball of $\rho$ is 
    \begin{align}
        \label{eqn:probability_larger_set_inside}
        & \Pr[\exists \mathcal{S}, |\mathcal{S}|\geq c+1, \forall i\in\mathcal{S}, \Delta(\hat{\rho}_i,\rho)\leq \epsilon] \\
        & = 1- \Pr[\exists \mathcal{R}, |\mathcal{R}|\geq c, \forall i\in\mathcal{R}, \Delta(\hat{\rho}_i,\rho)> \epsilon].
    \end{align}
    For every such set $\mathcal{R}$, we find in particular a subset $|\mathcal{R}'|=c$ with the same property. Therefore it is enough to bound the probability of finding such a subset, and we get
    \begin{align}    
        \text{Eq. \eqref{eqn:probability_larger_set_inside}} & \geq 1- \sum_{|\mathcal{R}'|=c} \Pr[\forall i\in\mathcal{R}', \Delta(\hat{\rho}_i,\rho)> \epsilon]\\
        & \geq 1- \binom{2c}{c} \xi^c \geq 1- (4\xi)^c. 
    \end{align}

    Second, if at least $c+1$ estimations are inside the $\epsilon$-ball of $\rho$, then $\hat{\rho}_{i^*}$ which maximizes the number of $\rho_j$ inside its $2\epsilon$-ball must be inside the $3\epsilon$ ball of $\rho$. The reason is as follows. In this case, $\Delta(\hat{\rho}_i,\hat{\rho}_j)$ between any $\hat{\rho}_i$ and $\hat{\rho}_j$ of the $c+1$ estimations inside the $\epsilon$-ball of $\rho$ is at most $2\epsilon$ by the triangle inequality. Therefore, the number of $\hat{\rho}_{j}$ inside the $2\epsilon$-ball of $\rho_{i^*}$ is at least $c$. On the contrary, $\Delta(\hat{\rho}_i,\hat{\rho}_k)\geq 2\epsilon$ for any $\hat{\rho}_i$ inside the $\epsilon$-ball of $\rho$ and any $\hat{\rho}_k$ outside the $3\epsilon$-ball of $\rho$ by the triangle inequality. As a result, the number of $\hat{\rho}_j$ inside the $2\epsilon$-ball of $\hat{\rho}_{k}$ for any $\hat{\rho}_k$ outside the $3\epsilon$-ball of $\rho$ is at most $c-1$. Therefore, $\rho_{i^*}$ which maximizes the number of $\hat{\rho}_j$ inside its $2\epsilon$-ball must be inside the $3\epsilon$-ball of $\rho$. 

    Combining both parts, we conclude that $\hat{\rho}_{i^*}$ obtained by maximizing the number of $\hat{\rho}_{j}$ inside its $2\epsilon$-ball is also inside the $3\epsilon$-ball of $\rho$ with a probability exponentially close to $1$, 
    \begin{align}
        \Pr[\Delta(\hat{\rho}_{i^*},\rho)\leq 3\epsilon]\geq 1-(4\xi)^c, 
    \end{align}
    which concludes our proof. 
\end{proof}

Based on Lemma~\ref{lem:pac_repetition}, we can prove the equivalence between the PAC condition and the expected squared distance condition up to logarithmic factors. 

\begin{lemma}
    \label{lem:cond_equiv}
    An algorithm satisfying the PAC condition can be converted into an algorithm satisfying the SMD condition with $O(\ln\frac{1}{\epsilon})$ repetitions, and an algorithm satisfying the SMD condition can be converted into an algorithm satisfying the PAC condition with $O(\ln\frac{1}{\xi})$ repetitions. 
\end{lemma}

\begin{proof}
    On the one hand, suppose that an algorithm fulfills the PAC condition, 
    \begin{align}
        \Pr[\Delta(\rho,\hat{\rho})\leq \epsilon] \geq 1-\xi. 
    \end{align}
    Using Lemma~\ref{lem:pac_repetition}, $\frac{4\ln 1/\epsilon}{\ln 1/4\xi }$ repetitions of the algorithm achieve 
    \begin{align}
        \Pr[\Delta(\rho,\hat{\rho})\leq 3\epsilon] \geq 1-\epsilon^2. 
    \end{align}
    It thus implies 
    \begin{align}
        & \expect[\Delta(\rho,\hat{\rho})^2]\nonumber \\
        & \leq 9\epsilon^2 \Pr[\Delta(\rho,\hat{\rho})\leq 3\epsilon] +  \Pr[\Delta(\rho,\hat{\rho}) > 3\epsilon]\leq 10\epsilon^2. 
    \end{align}
    As a result, the PAC condition implies the expected squared distance condition with at most a logarithmic factor $O(\ln\frac{1}{\epsilon})$ of repetitions.  
    
    On the other hand, suppose that an algorithm fulfills the expected squared distance condition, 
    \begin{align}
        \sqrt{\expect[\Delta(\rho,\hat{\rho})^2]}\leq \epsilon.
    \end{align}
    By the convexity of $f(x)=x^2$ we have
    \begin{align}
        \expect[\Delta(\rho,\hat{\rho})]  & \leq \epsilon, \\
        \variance[\Delta(\rho,\hat{\rho})]  & \leq \epsilon^2. 
    \end{align}
    By Chebyshev's inequality, 
    \begin{align}
        \Pr[\Delta(\hat{\rho},\rho)\leq 5\epsilon]\geq \frac{15}{16}. 
    \end{align}
    Using Lemma~\ref{lem:pac_repetition}, $\frac{\ln 1/\xi}{\ln 2}$ repetitions of the algorithm achieve 
    \begin{align}
        \Pr[\Delta(\hat{\rho},\rho)\leq 15\epsilon]\geq 1-\xi. 
    \end{align}
    Therefore, the expected squared distance condition implies the PAC condition with at most a logarithmic factor $O(\ln\frac{1}{\xi})$ of repetitions.
\end{proof}

\section{Algorithmic implementation}
The unitary Schur sampling and the discrete POVM measurement cannot be realized \emph{exactly} but \emph{approximately} with elementary one- and two-qubit gates. In this section, we discuss what accuracy the implementation should achieve.

Firstly, we restate Lemma \ref{lem:stream_wss} in a more precise way:

\begin{lemma}
    \label{lem:stream_wss_restated}
    Let $\sigma= \rho^{\otimes n}$ be an $n$-qudit state. 
    Algorithm~\ref{alg:stream_wss} outputs label $\lambda\vdash d$ and state $ \tilde{\sigma}_\lambda^n$ with probability $\tilde{p}_\lambda^n$, where 
   \begin{align}
       \tilde{p}_\lambda^n & = \Tr[(\mathbbm{I}_{\mathcal{P}_\lambda}\otimes \mathbbm{I}_{\mathcal{Q}_\lambda^d}) \tilde{U}_{\rm Sch}^n \sigma  \tilde{U}_{\rm Sch}^{n,\dagger}],\\
       \tilde{\sigma}_\lambda^n & = \frac{1}{\tilde{p}_\lambda^n} \Tr_{\mathcal{P}_\lambda}[(\mathbbm{I}_{\mathcal{P}_\lambda}\otimes \mathbbm{I}_{\mathcal{Q}_\lambda^d}) \tilde{U}_{\rm Sch}^n \sigma  \tilde{U}_{\rm Sch}^{n,\dagger}],
   \end{align}
   and $\|\tilde{U}_{\rm Sch}^n -U_{\rm Sch}^n \|_\infty\leq \xi$. When $\xi=0$, $\tilde{\sigma}_{\lambda}^n= \frac{1}{s_\lambda(\rho)}{\bm q}_\lambda(\rho)$ and $\tilde{p}_\lambda^n =\Tr[\sigma \Pi^n_\lambda]$. 
\end{lemma}

Next, we show:

\begin{lemma}
    \label{lemma:approximate_implementation_measurement}
    Let the failure probability  of the exact tomography algorithm be $\xi$. If the approximate tomography algorithm resembles the exact tomography algorithm with an accuracy $\xi$ in the operator norm, then the failure probability of the approximate tomography algorithm achieves $3\xi$, which can be easily resolved by a constant number of repetitions given $\xi<\frac{1}{12}$ due to Lemma~\ref{lem:pac_repetition}. 
\end{lemma}
\begin{proof}
    Now let the state to be estimated be $\rho$, the input be $\sigma$ and the output be $\hat{\rho}$. Let $U$ be the exact unitary we want to apply. Let $P(\hat{\rho})$ be the projection associated with the measurement outcome $\hat{\rho}$. Let $p_{\rm G}^\epsilon$ be the probability that $\Delta(\hat{\rho},\rho)\leq \epsilon$ for the exact tomography algorithm. Because the exact tomography algorithm satisfies the PAC condition, we have 
    \begin{align}
        p_{\rm G}^\epsilon = \sum_{\hat{\rho}:\Delta(\hat{\rho},\rho)\leq \epsilon} \Tr(P(\hat{\rho})U\sigma U^\dagger) \geq 1-\xi. 
    \end{align}
    Now let $\tilde{U}$ be the approximate unitary we can apply and suppose that $\|U-\tilde{U}\|_\infty\leq \xi$. Let $\tilde{p}_{\rm G}^\epsilon$ be the probability that $\Delta(\hat{\rho},\rho)\leq \epsilon$ for the approximate tomography algorithm. 
    \begin{align}
        \tilde{p}_{\rm G}^\epsilon = \sum_{\hat{\rho}:\Delta(\hat{\rho},\rho)\leq \epsilon}\Tr(P(\hat{\rho}) \tilde{U} \sigma \tilde{U}^\dagger).
    \end{align}
    Therefore, 
    \begin{align}
        p_{\rm G}^\epsilon - \tilde{p}_{\rm G}^\epsilon =  \Tr(\sum_{\hat{\rho}:\Delta(\hat{\rho},\rho)\leq \epsilon}P(\hat{\rho})(U\sigma U^\dagger - \tilde{U}\sigma \tilde{U}^\dagger) ),
    \end{align}
    and thus 
    \begin{align}
        & |p_{\rm G}^\epsilon - \tilde{p}_{\rm G}^\epsilon|  \leq  \Tr(\sum_{\hat{\rho}:\Delta(\hat{\rho},\rho)\leq \epsilon}P(\hat{\rho})|U\sigma U^\dagger - \tilde{U}\sigma \tilde{U}^\dagger| ) \\
        & \leq \Tr(|U\sigma U^\dagger - \tilde{U}\sigma \tilde{U}^\dagger|) = \|U\sigma U^\dagger - \tilde{U}\sigma \tilde{U}^\dagger\|_1. 
    \end{align}
    By the triangle inequality, 
    \begin{align}
        \|U\sigma U^\dagger - \tilde{U}\sigma \tilde{U}^\dagger\|_1 \leq 2\|(U  - \tilde{U})\sigma\|_1. 
    \end{align}
    By Holder's inequality, 
    \begin{align}
        \|(U  - \tilde{U})\sigma\|_1 \leq \|U-\tilde{U}\|_\infty \|\sigma\|_1 \leq \xi.
    \end{align}
    Combining all the inequalities, we obtain
    \begin{align}
        |p_{\rm G}^\epsilon - \tilde{p}_{\rm G}^\epsilon| \leq 2\xi, 
    \end{align}
    and therefore 
    \begin{align}
        \tilde{p}_{\rm G}^\epsilon \geq p_{\rm G}^\epsilon - 2\xi \geq 1-3\xi, 
    \end{align}
    which concludes our proof. 
\end{proof}

\section{Measuring POVM}
\label{apd:measure_povm}

In this appendix, we discuss the method to measure a POVM as well as its space and time complexity. From a high level, we first use the Naimark dilation to convert a POVM into an isometry and subsequently into a unitary, whose dimension upper bounds the memory complexity. We then use the Solovay-Kitaev theorem to upper bound the time complexity. 
\begin{lemma}
    \label{lem:povm_unitary}
    Let $\mathcal{H}$ be a finite dimensional Hilbert space with $\dim\mathcal{H}=D$, and let $\{E_x\}_{x\in X}$ be some POVM on $\mathcal{H}$ with $|X|=K$. Then we can construct a Hilbert space $\mathcal{H}_X$ with $\dim(\mathcal{H}_X)=K$ so that $\{E_x\}_{x\in X}$ is implementable on $\mathcal{H}\otimes\mathcal{H}_X$ by performing projective measurements on the computational basis of $\mathcal{H}_X$. This implementation uses $O(D^2K^2\ln\frac{DK}{\xi})$ elementary gates and $O(\ln K)$ additional qubits of memory, where $\xi$ is the error of the unitary implementing the measurement.
\end{lemma}

\begin{proof}
    Let $\{\ket{x}\}_{x\in X}$ be the computational basis on $\mathcal{H}_X$ and let $\{\ket{d}\}_{d\in[D]}$ be the computational basis of $\mathcal{H}$. We take the Naimark dilation of the measurement~\cite[Theorem 4.6]{Paulsen_2003} given by $V:\mathcal{H}\to\mathcal{H}\otimes\mathcal{H}_X$ defined as
    \begin{align}
        V:=\sum_{x\in X}\sqrt{E_x}\otimes\ket{x}.
    \end{align}
    We now want to extend $V$ to a unitary  $U:\mathcal{H}\otimes\mathcal{H}_X\to\mathcal{H}\otimes\mathcal{H}_X$. For this, we find an orthonormal basis of $\text{ran}(V)^{\perp}$. Since $V$ is an isometry, $\dim(\text{ran}(V))=D$, and thus $\dim(\text{ran}(V)^{\perp})=D(K-1)$. Therefore we can label the new basis vectors by $\{\ket{U_{d,x}}\}_{d\in [D], x\neq x_{0}}$, where the labelling is arbitrary, and $x_{0}\in X$ is arbitrary as well. Now we can define
    \begin{align}
        U:=\sum_{x\in X}\sqrt{E_x}\otimes\ketbra{x}{x_{0}}+\sum_{x\neq x_{0}\in X}\sum_{d=1}^D\ketbra{U_{d,x}}{d,x}.
    \end{align}

    Initializing System $\mathcal{H}_{X}$ in the state $\ket{x_{0}}$, then performing $U$ and afterwards performing the PVM $\{\id_{\mathcal{H}}\otimes\ketbra{x}{x}\}_{x\in X}$ now implements the POVM $\{E_x\}_{x\in X}$.

    The ancilla memory complexity is now $O(\ln K)$, because we need at most $O(\ln K)$ extra qubits to store $\mathcal{H}_X$ with $\dim\mathcal{H}_X = K$. 
    
    It is now a well known fact (see e.g. \cite{Shende_2006,Iten_2016}) that we can implement $U$ using $O(\dim(\mathcal{H}\otimes\mathcal{H}_X)^2)=O(D^2K^2)$ one-qubit gates and CNOT gates. Using the Solovay-Kitaev algorithm~\cite{Nielsen_2010,Selinger_2015}, we can in turn implement each one-qubit gate in $O(\ln\frac{DK}{\xi})$ gates to achieve an overall operator norm error of $\xi$. In all, this gives $O(D^2K^2\ln\frac{DK}{\xi})$ elementary gates to implement the measurement, where the measurement unitary has error at most $\xi$.
\end{proof}

\section{Reconstruction map}
\label{apd:reconstruction}

In this section, we remark that provided that the sampled set $\mathcal{S}$ obeys Definition~\ref{def:property_S} and~\ref{def:property_T}, it can be used to perform the required POVM and reconstruct the unknown state for the fidelity and trace distance cases for any input state $\rho$. This justifies our choice of forgoing the complexity analysis of sampling $\mathcal{S}$ in the analysis of the tomography algorithm since the POVM derived from $\mathcal{S}$ may be reused in independent applications of our tomography algorithm. 

We show that the choice of $\mathcal{S}$ is independent of $\rho$ in a more explicit way here. As we have shown in Lemma~\ref{lem:discrete_povm}, 
\begin{align}
    (1-\eta)\mathcal{W}_\lambda\leq \mathcal{W}_\lambda^\mathcal{S} \leq (1+\eta)\mathcal{W}_\lambda,
\end{align}
which is independent of the choice of $\rho$. Now we consider the reconstruction maps which take input on system $A$ and produce output on system $B$ before and after discretisation. The reconstruction maps before and after discretisation both start with the unitary Schur sampling, in which we obtain $\lambda$ and $\rho_\lambda=\frac{1}{s_\lambda(\rho)}{\bm q}_\lambda(\rho)$. Therefore, we only need to compare the reconstruction maps before and after discretisation restricted to this $\lambda$-subspace. For any $\rho$, the reconstruction maps restricted to the $\lambda$-subspace which maps $\rho^{\otimes n}$ to an estimate $\hat{\rho}$ before and after discretisation are, respectively
\begin{align}
    R_\lambda(\rho_\lambda) & = \frac{\dim \mathcal{Q}_\lambda^d}{s_\lambda(\overline{\lambda})} \int d U \Tr({\bm q}_\lambda(U\overline{\lambda}U^\dagger) \rho_\lambda ) \otimes U\overline{\lambda}U^\dagger \\
    & = \frac{\dim \mathcal{Q}_\lambda^d}{s_\lambda(\overline{\lambda})} \Tr_A\left((\sqrt{\rho_\lambda} \otimes \id_B) \mathcal{W}_\lambda (\sqrt{\rho_\lambda} \otimes \id_B)  \right) , \\
    R_\lambda'(\rho_\lambda) & = \frac{\dim \mathcal{Q}_\lambda^d}{(1+\eta)|\mathcal{S}|s_\lambda(\overline{\lambda})}\sum_{U\in\mathcal{S}} \Tr({\bm q}_\lambda(U\overline{\lambda}U^\dagger) \rho_\lambda) \otimes U\overline{\lambda}U^\dagger \\
    & = \frac{\dim \mathcal{Q}_\lambda^d}{(1+\eta)s_\lambda(\overline{\lambda})} \Tr_A\left((\sqrt{\rho_\lambda} \otimes \id_B ) \mathcal{W}_\lambda^\mathcal{S} (\sqrt{\rho_\lambda} \otimes \id_B) \right) . 
\end{align}
By Lemma~\ref{lem:discrete_povm}, this leads to the fact for any $\rho_\lambda$, it holds that 
\begin{align}
    \frac{1-\eta}{1+\eta}R_\lambda(\rho_\lambda) \leq R_\lambda'(\rho_\lambda) \leq  R_\lambda(\rho_\lambda), 
\end{align}
which means that no matter what the input state is, the reconstruction map after discretisation works equally well as the reconstruction map after the discretisation: the reconstructed states using either map are close in relative error. This equivalence holds both for the infidelity and trace distance case.

\section{Recursive POVM}
\label{apd:recuresive_povm}
In this appendix we introduce a recursive algorithm to implement a large POVM in a memory-efficient way. It can help to reduce the memory complexity of any arbitrary POVM to a constant number of auxiliary qubits. This is due to the fact that in each step, we have just $k$ outcomes, with $k$ arbitrary. However, as the size of the POVM in Algorithm \ref{alg:stream_tomo} is not too large, this is not necessary for our main result.

\begin{figure}[htbp!]
\begin{algorithm}[H]\label{alg:recursive_povm}
\SetAlgoLined
\caption{\textbf{RecursivePOVM}}
\SetKwInOut{Input}{Input}
\SetKwInOut{Output}{Output}
\SetKwInOut{Init}{Initialization}
\SetKwRepeat{Repeat}{repeat}{until}
\Input{State $\rho$ over Hilbert space $\mathcal{H}$\\
POVM $\{E_x\}_{x\in\mathcal{X}}$ over Hilbert space $\mathcal{H}$\\
Number of partitions $k$}
\Init

\eIf{
$\frac{|\mathcal{X}|}{k}\leq 1$
}{
Measure with POVM $\Ex$ to receive $x\in\mathcal{X}$\;
\Return $x\in\mathcal{X}$
}{
Partition $\mathcal{X}=\bigsqcup_{i=1}^k\mathcal{X}_i$\;
Set $G_i=\sum_{x\in\mathcal{X}_i}E_x$\;
Measure with POVM $\Gi$ to receive $i\in[k]$ and post-measurement state $\tilde{\rho}^i=\frac{G_i^{\frac{1}{2}}\rho G_i^{\frac{1}{2}}}{\Tr[\rho G_i]}$\;
Set $F_x^i=G_i^{-\frac{1}{2}}E_xG_i^{-\frac{1}{2}}$ for all $x \in\mathcal{X}_i$\;
Set $F^{i}_{\perp}=\id_{\mathcal{H}}-\sum_{x \in\mathcal{X}_i}F_x^i$\;
\Return \textbf{RecursivePOVM}$\big(\tilde{\rho}^i, \Fxi\cup\{F^{i}_{\perp}\}, k\big)$
}
\end{algorithm}
\end{figure}

The operators $G_i^{-\frac{1}{2}}$ used in this algorithm are the pseudo-inverses for the operators $G_i^{\frac{1}{2}}$. The following Lemma proves the operators $\Fxi\cup\{F^{i}_{\perp}\}$ defined in Algorithm~\ref{alg:recursive_povm} indeed form a POVM.

\begin{lemma}
    The operators $\Fxi\cup\{F^{i}_{\perp}\}$ defined in Algorithm \ref{alg:recursive_povm} form a POVM.
\end{lemma}

\begin{proof}
    We have
    \begin{align}
        \sum_{x \in\mathcal{X}_i}F_x^i=\sum_{x \in\mathcal{X}_i}G_i^{-\frac{1}{2}}E_xG_i^{-\frac{1}{2}}=G_i^{-\frac{1}{2}}G_iG_i^{-\frac{1}{2}}=P_{i}.
    \end{align}
    Here, $P_{i}$ denotes the projector onto the range of $G_{i}$. Since $F^{i}_{\perp}=\id_{\mathcal{H}}-\sum_{x \in\mathcal{X}_i}F_x^i$, we find that $0\preceq F^{i}_{\perp}$. Further, since $G_{i}$ is a positive sum of positive operators, it is again positive, and therefore $G_i^{-\frac{1}{2}}$ is positive as well. This means we have $0\preceq G_i^{-\frac{1}{2}}E_x G_i^{-\frac{1}{2}}=F_x^i$. Therefore, all operators are positive. Finally, we can see from the definition that
    \begin{align}
        \id_{\mathcal{H}}=F^{i}_{\perp}+\sum_{x \in\mathcal{X}_i}F_x^i .
    \end{align}
\end{proof}

To show correctness of Algorithm \ref{alg:recursive_povm}, we want to prove that if the initial call receives the state $\rho$ and the POVM $\{E_x\}_{x\in\mathcal{X}}$ then its final output $x$ is produced with the same probability as that given by Born's rule: $\Tr[E_x\rho]$.
Formally:
\begin{lemma}\label{lem:rec_povm}
    Let $X$ be the random variable over alphabet $\mathcal{X}$ denoting the final output of Algorithm~\ref{alg:recursive_povm} on initial input of state $\rho$ and POVM $\Ex$.
    We have
    \begin{align}
        \Tr[E_x\rho] = \Pr[X=x]
    \end{align}
    for all $x\in\mathcal{X}$.
\end{lemma}

\begin{proof}
    We will prove the correctness inductively. To start, if $|\mathcal{X}|\leq k$ we see that
    \begin{align}
        \Tr[E_x\rho] = \Pr[X=x]
    \end{align}
    for all $x\in \mathcal{X}$.
    
    Now we take $\mathcal{X}=\bigsqcup_{i=1}^k\mathcal{X}_i$ and we denote
    \begin{align}
        \Tr[F^i_x\tilde{\rho}^i] = \Pr[X=x|I=i] , \\
        \Tr[G_i\rho] = \Pr[I=i] .
    \end{align}
    for all $i\in[k]$ and $x\in \mathcal{X}_i$. It holds that
    \begin{align}
        \Pr[X=x]=\Pr[X=x|I=i]\Pr[I=i] .
    \end{align}
    Inserting the equations above and the definition of $\tilde{\rho}^i$ gives
    \begin{align}
        \Pr[X=x]=\Tr[G_{i_x}^{\frac{1}{2}}F^{i_x}_xG_{i_x}^{\frac{1}{2}}\rho] .
    \end{align}
    We remember that $F_x^{i_x}=G_{i_x}^{-\frac{1}{2}}E_xG_{i_x}^{-\frac{1}{2}}$, where $G_{i_x}^{-\frac{1}{2}}$ is the pseudo-inverse of $G_{i_x}^{\frac{1}{2}}$. This means we have
    \begin{align}
        G_{i_x}^{\frac{1}{2}}F^{i_x}_xG_{i_x}^{\frac{1}{2}}=P_{i_x}E_xP_{i_x}=E_x .
    \end{align}
    Here, $P_{i_x}$ is a projection onto range of $G_{i_x}$. Since $G_i=\sum_{y\in\mathcal{X}_i}E_y$ and all $E_y$ are positive, we have $\ran{E_x}\subseteq\ran{G_{i_x}}$, which gives the last equality. Inserting into the equations above, we get
    \begin{align}
        \Pr[X=x]=\Tr[G_{i_x}^{\frac{1}{2}}F^{i_x}_xG_{i_x}^{\frac{1}{2}}\rho]=\Tr[E_x\rho]  .
    \end{align}
    This completes the induction step. 
\end{proof}

\nocite{*}

\end{document}